\documentclass[letterpaper,prb,superscriptaddress,showpacs,twocolumn]{revtex4}

\usepackage{graphicx}
\usepackage{amsmath}
\newcommand{\bs}[1]{\boldsymbol{#1}}
\newcommand{\half}{\textstyle{\frac{1}{2}}}
\newcommand{\sfrac}[2]{{\textstyle{\frac{#1}{#2}}}}

\begin{document}

\title{Phase diagrams of the metallic zigzag carbon nanotube}
\author{J. E. Bunder}
\affiliation{Department of Physics, National Tsing-Hua University, Hsinchu 300, Taiwan}
\affiliation{Physics Division, National Center for Theoretical Sciences, Hsinchu 300, Taiwan}
\author{Hsiu-Hau Lin}
\affiliation{Department of Physics, National Tsing-Hua University, Hsinchu 300, Taiwan}
\affiliation{Physics Division, National Center for Theoretical Sciences, Hsinchu 300, Taiwan}
\date{\today}

\begin{abstract}

We investigate a metallic zigzag carbon nanotube by means of a Hubbard model which includes both on-site and nearest neighbour interactions. Assuming weak interactions, a renormalization group analysis of the equivalent two-leg ladder followed by bosonization and refermionization results in a Gross-Neveu model with an enlarged symmetry relative to the original Hamiltonian. For the undoped case the symmetry of the Gross-Neveu model is SO(8), but for the doped case the particle-hole symmetry is broken and the symmetry reduces to SO(6). Four ground state phases are found in the undoped carbon nanotube with repulsive interactions, a $d$-wave Mott insulator, an $s$-wave Mott insulator, a $p$-density wave and a charge density wave. The doped case has two ground state phases, a $d$-wave superconductor and a phase where a $p$-density wave and a charge density wave co-exist. We also explore the global phase diagram with a general interaction profile and find several additional states, including a chiral current phase where current flows around the nanotube along the zigzag bonds.   
\end{abstract}

%05.10.Cc Renormalization group methods  
%61.50.Ah Theory of crystal structure, crystal symmetry; calculations and modeling  
%61.72.Ww Doping and impurity implantation in other materials  
%64.60.Ak Renormalization-group, fractal, and percolation studies of phase transitions  
%71.10.Fd Lattice fermion models (Hubbard model, etc.)  
%71.10.Hf Non-Fermi-liquid ground states, electron phase diagrams and phase transitions in model systems  
%73.22.-f Electronic structure of nanoscale materials: clusters, nanoparticles, nanotubes, and nanocrystals  

\maketitle

\section{introduction}
\label{intro}
Carbon nanotubes (CNT) are long, thin tubes constructed from sheets of graphite. Because of their many novel properties CNT have numerous potential applications in material science,~\cite{Dalton03,Zhang05,Koziol07} optics~\cite{Zhu06,Yang08} and electronics,\cite{Tans98,Novak04,Sternberg06} while also contributing to our knowledge of fundamental physics.~\cite{Romero05,Dresselhaus07,Meyer07} They are extremely strong, owing to being constructed from $sp^2$ bonds, and yet their low density makes them extremely light. CNT have the highest tensile strength and elastic modulus of any known material~\cite{Pan99,Yu00} but this impressive strength is only applicable to forces which stretch the nanotube as their hollow structure means they readily become distorted under torsion, compression or bending.~\cite{Yakobson96} The electrical properties of a CNT depend on its structure. Single walled CNT are generally classified as one of three types: zigzag, armchair or chiral, and these classifications are defined by the orientation of the graphene lattice about the tube. From band structure calculations, armchair CNT are always metallic with current densities which may possibly exceed silver and copper,~\cite{Yao00} while zigzag and chiral CNT can be metallic, semiconducting or insulating depending on the width and helicity of the tube.~\cite{Mintmire92,Hamada92,Saito92,Odom98,Wilder98} 

The simple band structure calculations used to determine whether or not a CNT is metallic are performed in the weakly interacting limit. In general this limit is not applicable to CNT as they tend to have long-range Coulomb interactions which are not small, though one can justify considering only weak short-range interactions if the CNT is screened. Screening can be achieved either by arranging several nanotubes in an array or rope-like structure,~\cite{Gonzalez05,Gonzalez06} or by placing a single nanotube close to a conducting plate.~\cite{Hausler02,Fogler05}  
Studies on armchair CNT with long-range interactions have shown that the doped CNT ground state is a metallic Tomanaga-Luttinger liquid~\cite{Kane97,Egger97} while the ground state of the undoped CNT has a number of possible phases including Mott insulators and density waves.~\cite{Yoshioka99,Nersesyan03} 
Some of these phases have also been found in doped and undoped armchair CNT with short-range interactions.~\cite{Lin98,Konik00}

Here we consider a metallic zigzag CNT which we assume to be screened so that only on-site and nearest neighbour interactions need to be considered. Our main purpose is to determine which ground state phases can be supported by a metallic zigzag CNT with short-range interactions, and to describe the phase transitions between these phases. The Hubbard Hamiltonian of a metallic CNT may be mapped onto the well known Hubbard Hamiltonian of a two-leg ladder,~\cite{Balents97,Krotov97,Lin98a} though the nature of the two-leg ladder is dependent on the chirality of the nanotube.
The armchair CNT maps onto a fairly standard form of the two-leg ladder where hopping between any two adjacent lattice sites along either leg is always the same, as is hopping along any rung i.e., between legs. So, it is not hard to determine the behaviour of an armchair CNT directly from known results of two-leg ladders.~\cite{Lin98,Konik00,Krotov97,Tsuchiizu02,Fjaerestad02,Momoi05,Tsuchiizu05} In contrast, the metallic zigzag CNT maps onto an unusual type of two-leg ladder in which the hopping part describes two chains with hopping strength alternating between lattice sites, but no hopping between legs. The two chains however cannot be described as independent as they influence each other through Coulomb interactions. 

The nearest neighbour Coulomb interactions in an armchair CNT map rather simply onto a two-leg ladder, acting between nearest neighbours either along the rungs or along the legs.~\cite{Lin98a} The situation is quite different in the two-leg ladder equivalent of a zigzag CNT. In this case the interactions in the ladder act either between nearest neighbours along the legs or between next nearest neighbours on different legs. Initially we hoped that these more complex interactions would have an interesting effect on the phase diagram, possibly allowing some unusual phases with broken time-reverasl symmetry. For instance, it is interesting to explore whether a staggered flux phase exists under physically possible conditions. As we shall show later, this is not the case and the physically possible phases of a zigzag CNT are similar to those found in an armchair CNT. 

\begin{figure}
\includegraphics[width=8cm]{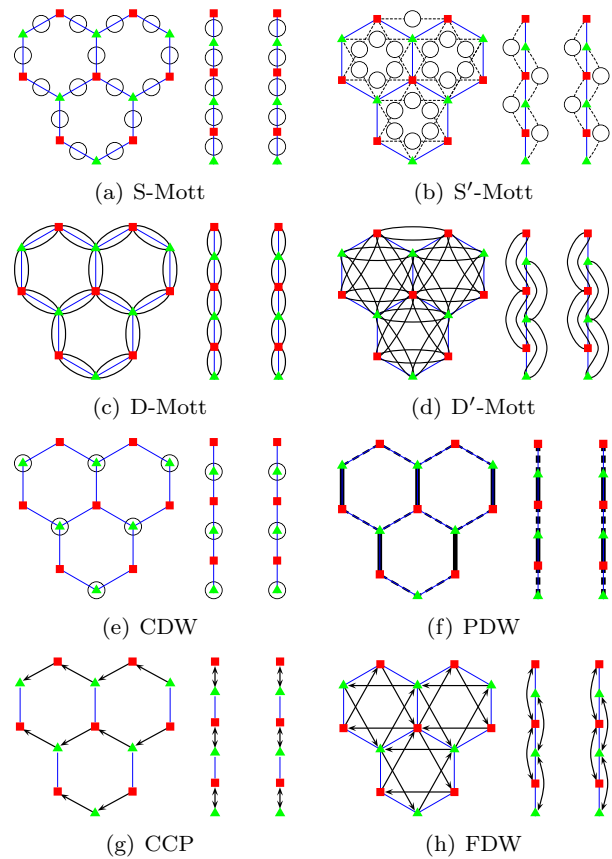}
\caption{Eight phases of the zigzag CNT and the equivalent two-leg ladder in the strong coupling limit. The symbols are defined in the text.}
\label{fig:phases}
\end{figure}

We determine that our metallic CNT can support at least seven different phases when there is no doping, though some of these phases are only found when attractive interactions are permitted. Four phases are Mott insulators, two with $d$-wave symmetry which we name D-Mott and D$'$-Mott, and two with $s$-wave symmetry which we name S-Mott and S$'$-Mott. These four Mott phases are analogous to those found in a standard two-leg ladder and so we have adopted the same naming convention.~\cite{Tsuchiizu02}
The oder parameters of currents and/or bond hopping are always zero
in the Mott insulator phases, and the average electron density is always one electron per site. These phases are classified as $d$-wave or $s$-wave based on the nature of the pairing correlations. Illustrations of all four Mott states in both the honeycomb lattice of the zigzag CNT and the equivalent two-leg ladder are shown in Figs. \ref{fig:phases}(a),(b),(c) and (d). In these figures the circles represent an $s$-wave pairing of two electrons with opposite spin. One pair must occupy one of two possible sites but which site is chosen is completely random, ensuring an average electron density of one electron per site. In the S-Mott the two possible positions for one electron pair are nearest neighbours so we draw circles between all nearest neighbours. In the S$'$-Mott the two possible positions for one pair are next-nearest neighbours so the circles are drawn between next-nearest neighbours. The $d$-wave pairing is represented by ellipses. In this case paired electron occupy different sites, though the positions of the different spins are not ordered but random. The paired electrons are nearest neighbours in the D-Mott so the ellipses are drawn along the bonds joining nearest neighbours. In the D$'$-Mott the pairing is between next-nearest neighbours so the ellipses (distorted for clarity in the two-leg ladder case) join next-nearest neighbour sites.

The remaining three phases are density waves which all exhibit a broken Z$_2$ symmetry. The charge density wave (CDW) has broken particle-hole symmetry so there is a regular variation in the electron distribution along the lattice as shown in Fig. \ref{fig:phases}(e). As before the circles represent a pair of opposite spin electrons. A major difference between the CDW and the two $s$-wave Mott states is that positions of the pairs in the CDW is not random. The $p$-density wave (PDW) is equivalent to a spin-Peierls state where dimers form between neighbouring sites as shown in Fig. \ref{fig:phases}(f). The thickness of the dashed and solid black lines indicates the magnitude of the kinetic energy exchange between sites, though these two types lines have opposite signs.  The chiral current phase (CCP) describes a state in which a current circulates around the nanotube, flowing between nearest neighbours. This is represented in Fig. \ref{fig:phases}(g) with the arrows describing the current. In the equivalent two-leg ladder all currents vanish because the two-leg ladder lies along the longitudinal axis of CNT and the net current along the longitudinal axis of the CNT is zero. There is fourth density wave phase which is theoretically possibly, though we do not find it in our phase diagram. This phase is a $f$-density wave (FDW) which has a circulating current which flows between next-nearest neighbours, as shown in Fig. \ref{fig:phases}(h). The current in the equivalent two-leg ladder vanishes, again because the net current along the CNT longitudinal axis is zero.

The eight phases are essentially equivalent to phases found in the armchair CNT (though the phase diagram is not identical), and comparable to phases found in the two-leg ladder, with the exception of the CCP. When an armchair CNT is in a CCP the equivalent two-leg ladder is in a staggered flux (SF) phase in which current flows around plaquettes between nearest neighbours, with the direction of the current being opposite in neighbouring plaquettes.~\cite{Fjaerestad02} The two-leg ladder equivalent of the zigzag CNT's CCP resembles a SF phase, yet it cannot be a true SF phase as this ladder does not have standard plaquettes around which a non-zero current may flow.

If a two-leg ladder or CNT is lightly doped away from half-filling different phases emerge, yet they are still closely related to the phases found in the half-filled case. Two phases are superconducting, one being $d$-wave (D-SC) and the other $s$-wave (S-SC). The D-SC can be thought of as a merging of the D-Mott and the D$'$-Mott insulator phases found in the undoped case. Similarly, the S-SC is a merging of the undoped S-Mott and S$'$-Mott phases. The doped two-leg ladder and CNT have two density wave phases. One is a combination of the undoped CDW and PDW phases. For simplicity we will refer to this doped phase as a CDW. The other density wave is a combination of the CCP and the FDW in the CNT (or a SF and FDW in the two-leg ladder), and we shall refer to this phase as a CCP. The general appearance of the phase diagrams of the doped and undoped cases are quite similar, although, in the undoped cases the Mott insulator phases tend to dominate most of the phase diagram, while on doping the density wave regions are significantly enhanced. 

One fascinating characteristic of undoped two-leg ladders, also shown to exist in CNT, is dynamical symmetry enlargement (DSE). In the two-leg ladder the symmetry of the original Hubbard Hamiltonian is U(1)$\times$SU(2) but the effective Hamiltonian obtained after solving the renormalization group (RG) equations is equivalent to a Gross-Neveu (GN) model with a SO(8) symmetry, which is far larger than the original symmetry.~\cite{Lin98} All phases exhibit DSE, yet they do not all share the same SO(8) symmetry. A doped two-leg ladder can still be mapped onto a GN model with DSE, but with the smaller SO(6) symmetry.~\cite{Tsuchiizu05} It is important to note that DSE is a consequence of weak coupling and begins to break down as the coupling increases.~\cite{Bunder07} Furthermore, the SO(8) symmetry is not particularly realistic as it depends on an equal charge and spin gap, while in reality the charge gap is substantially larger than the spin gap. In contrast, the SO(6) symmetry is fairly realistic as doping removes the charge gap but has no effect on the spin gap. 

In Sec. \ref{sec:model} we show how a Hubbard model describing a metallic zigzag CNT at half-filling with both on-site and nearest neighbour interactions may be mapped onto a two-leg ladder. The Hamiltonian is then bosonized. In Sec. \ref{sec:phases} we obtain all phases and phase transitions by solving the RG equations under different initial conditions and substituting the results into the bosonized Hamiltonian. We then show how the effective Hamiltonian can be represented as a SO(8) GN model with enlarged symmetry. We use a variety of order parameters such as the current and the electron density to determine the characteristics of each phase. We discuss the nature of the phase transitions  and show that they can mostly be classified as either Gaussian or Ising. In Sec. \ref{sec:doped} we consider the case of a doped CNT. The Hamiltonian is similar to the undoped Hamiltonian, except for the absence of Umklapp interactions. Using a RG analysis and bosonization we find the doped nanotube's ground state phases and the phase transitions between them and we show that all phases map to an effective SO(6) GN Hamiltonian. Finally, in Sec. \ref{sec:QN} we discuss some general properties of the SO(2N) GN model.

\section{the model}
\label{sec:model}

In this section we show how the zigzag CNT Hubbard model with on-site and nearest neighbour interactions may be mapped onto a two-leg ladder Hubbard model. After some standard approximations we bosonize the Hamiltonian. These derivations are not new,~\cite{Lin98,Lin98a} except for the addition of the nearest neighbour interactions, and therefore our derivation will be rather brief. 

\begin{figure}
\begin{center}
\includegraphics[width=7cm]{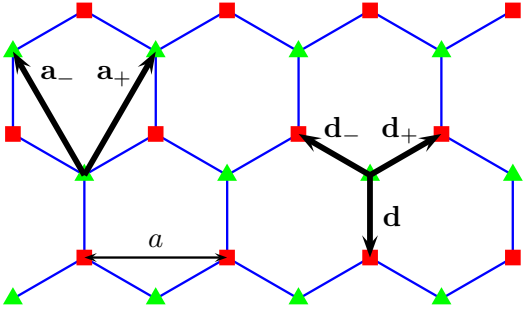}
\caption{The graphene lattice with the two triangular sublattices shown as red squares and green triangles.}
\label{fig:lattice}
\end{center}
\end{figure}

A carbon graphene lattice may be represented by two regular triangular sublattices, offset by $\mathbf{d}=a(0,-1/\sqrt{3})$ and with sublattice basis vectors $\mathbf{a}_{\pm}=a(\pm 1/2,\sqrt{3}/2)$, where $a$ is the sublattice lattice constant, as illustrated in Fig. \ref{fig:lattice}. 
The Hubbard model hopping Hamiltonian of such a carbon lattice is~\cite{Lin98a}
\begin{align}
H_0=&-t\sum_{\mathbf{r}=\mathbf{R},\alpha} [c_{1\alpha}^{\dag}(\mathbf{r})c_{2\alpha}
(\mathbf{r}+\mathbf{d}_+)
+c_{1\alpha}^{\dag}(\mathbf{r})c_{2\alpha}(\mathbf{r+d_-})]
\nonumber\\ 
&-t_{\perp}\sum_{\mathbf{r}=\mathbf{R},\alpha}
[c_{1\alpha}^{\dag}(\mathbf{r})c_{2\alpha}(\mathbf{r+d})+h.c]
\end{align}
where ${\bf d}_{\pm}={\bf a}_{\pm}+{\bf d}$, $\mathbf{R}=n_+\mathbf{a}_++n_-\mathbf{a}_-$ with integral $n_{\pm}$ describes a lattice vector in the first sublattice, $t_{\perp}$ is the hopping strength along the vertical (i.e., $y$ direction) bond and $t$ is the hopping strength along the other two bonds. In all calculations we define $t=t_{\perp}$. The annihilation operator $c_{i\alpha}$ describes the destruction of a fermion with spin $\alpha$ in the $i$th sublattice. 
The on-site interaction Hamiltonian is 
\begin{equation}
H_U=U\sum_{\mathbf{r}=\mathbf{R}',i}:n_{i\uparrow}n_{i\downarrow}:
\end{equation}
where $n_{i\alpha}=c^{\dag}_{i\alpha}c_{i\alpha}$, every lattice site is given by $\mathbf{R}'=n_+\mathbf{a}_++n_-\mathbf{a}_-+n_d\mathbf{d}$ with integral $n_{\pm,d}$, and $U$ is the on-site interaction strength. Similarly, the nearest neighbour interaction Hamiltonian is
\begin{align}
H_V=&V\sum_{\alpha\beta, {\bf r}\in {\bf R}}[n_{1\alpha}({\bf r}) n_{2\beta}({\bf r}+{\bf d}_{+})
+n_{1\alpha}({\bf r})n_{2\beta}({\bf r}+{\bf d}_{-})]
\nonumber\\
&+V_{\perp}\sum_{\alpha\beta, {\bf r}\in {\bf R}}
n_{1\alpha}({\bf r}) n_{2\beta}({\bf r}+{\bf d}) 
\end{align}
where $V_{\perp}$ is the nearest neighbour interaction strength across the vertical bond and $V$ is the nearest neighbour interaction strength across the other two bonds. We will only consider $|V|,|V_{\perp}|<|U|$.

A single walled CNT is formed by making a cylinder out of any graphene lattice, such as the one shown in Fig. \ref{fig:lattice}. If this lattice is rolled along a horizontal axis so that the top is joined to the bottom of the lattice (while retaining the regular hexagonal structure) we obtain an armchair CNT. If instead we roll this lattice along a vertical axis, joining the left and right sides, we obtain a zigzag CNT. A chiral CNT is any other cylinder which can be created from a graphene lattice which is neither armchair nor zigzag.     
Here we will only consider a metallic zigzag CNT. To determine when the zigzag CNT is metallic we derive the band structure in the weak coupling limit. 

The $y$-axis of the zigzag CNT is defined to be in the longitudinal direction of the nanotube and the $x$-axis is around the nanotube in the transverse plane.
As the momentum must be quantized in the $x$ direction 
\begin{equation}
k_x=\frac{2\pi p}{aN_x},\qquad p=0,\pm 1,\ldots,\pm(N_x/2),
\label{eq:kx}
\end{equation}
where $N_x$ is the number of lattice sites around the circumference of the CNT and the circumference is $aN_x$.
The CNT can be defined as metallic when at least one of these quantized momenta coincide with at least one of the Dirac points, defined as the zeros of the energy spectrum.
In the weak coupling limit $U,V,V_{\perp}\ll t,t_{\perp}$ the hopping Hamiltonian is sufficient for describing the band structure. In momentum space the energy spectrum obtained from the hopping Hamiltonian is $E(\mathbf{k})=\mp|h(\mathbf{k})|$ where
\begin{equation}
h(\mathbf{k})=2t\cos(k_xa/2)e^{ik_ya/2\sqrt{3}}
+t_{\perp}e^{-ik_ya/\sqrt{3}}.
\end{equation}
Therefore the Dirac points for $t=t_{\perp}$ are given by $\mathbf{k}=(\pm 4\pi/3a,0),(\pm 2\pi/3a,\pm 2\pi/\sqrt{3}a)$. The Dirac points which corresponds to one of the quantized momenta are
 $k_x=\pm 2\pi/3a$. This choice of quantized momenta restricts $N_x$ to being a multiple of three, i.e., a zigzag CNT is only metallic when $N_x$ is a multiple of three.

\begin{figure}
\begin{center}
\includegraphics[width=7cm]{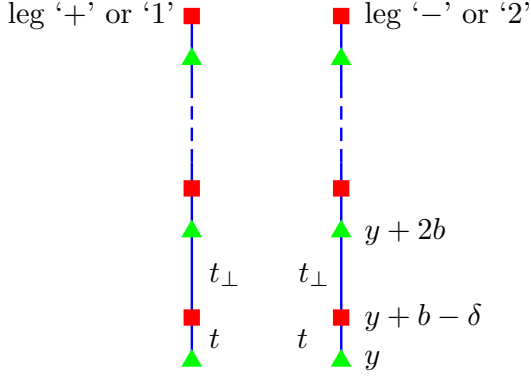}
\caption{A two-leg ladder which is equivalent to a metallic zigzag CNT with hopping $t,t_{\perp}$. We do not show the interactions here as they are a little complicated. The three values on the bottom right give general $y$ coordinates of the lattice points.}
\label{fig:2LLlattice}
\end{center}
\end{figure}

The Hubbard model of a metallic CNT which is either armchair or zigzag may be mapped onto a two-leg ladder after taking a Fourier transform of the transverse axis.~\cite{Balents97,Krotov97,Lin98a}
If the annihilation operator $c_i(\mathbf{r})$ is partially Fourier transformed in the $x$ direction and $k_x=\pm 2\pi/3a$,
\begin{equation}
c_i(x,y)=\sfrac{1}{\sqrt{N_x}}
\sum_{q=\pm}d_{qi}(y)e^{iq(2\pi/3a)x}.\label{eq:partialFT}
\end{equation}
On substitution of this Fourier transform into the hopping Hamiltonian,
\begin{align}
H_0=&\sum_{yq\alpha}[-td_{q1\alpha}^{\dag}(y)
d_{q2\alpha}(y+b_-)\nonumber\\
&-t_{\perp}d_{q1\alpha}^{\dag}(y)d_{q2\alpha}
(y-b_+)+h.c]\label{eq:hopH}
\end{align}
where $b_{\pm}=b\pm\delta$, $b=a\sqrt{3}/4$, $\delta=a/4\sqrt{3}$ and the Fermi point is $k_F=k_y=\pi/2b$. 
 Similarly, the two interaction Hamiltonians are
\begin{align}
H_U=&\sfrac{U}{N_x}\sum_{yqi}[n_{qi\uparrow}(y)n_{qi\downarrow}(y)
+n_{qi\uparrow}(y)n_{\bar{q}i\downarrow}(y)\nonumber\\
&+d^{\dag}_{qi\uparrow}(y)d_{\bar{q}i\uparrow}(y)
d^{\dag}_{\bar{q}i\downarrow}(y)d_{qi\downarrow}(y)]
\end{align}
with $n_{qi\alpha}=d^{\dag}_{qi\alpha}d_{qi\alpha}$ and $\bar{q}=-q$, and
\begin{multline}
H_V=\sfrac{2V}{N_x}\sum_{y q q'\alpha\beta} [n_{q1\alpha}(y)n_{q'2\beta}(y+b_-)\\
+\delta_{q'\bar{q}}\cos\sfrac{2\pi}{3}d_{q1\alpha}^{\dag}(y) d_{q'1\alpha}(y)d^{\dag}_{q'2\beta}(y+b_-) d_{q2\beta}(y+b_-)]\\
+\sfrac{V_{\perp}}{N_x}\sum_{\alpha\beta y q q'}[n_{q1\alpha}(y)n_{q'2\beta}(y-b_+)\\
+\delta_{q'\bar{q}}d_{q1\alpha}^{\dag}(y) d_{q'1\alpha}(y)d^{\dag}_{q'2\beta}(y-b_+) d_{q2\beta}(y-b_+)].\label{eq:nnH}
\end{multline}
Though the Hamiltonian now resembles a two-leg ladder Hamiltonian,  it is not in the standard form. The two-leg ladder Hamiltonian usually describes hopping and nearest neighbour interactions both along legs and across rungs, and this is the type of two-leg ladder the armchair CNT maps onto. However, the hopping part of the our Hamiltonian describes a two-leg ladder with no hopping across the rungs and a hopping strength alternating between $t$ and $t_{\perp}$ along the legs, as shown in Fig. \ref{fig:2LLlattice}. Note that although the two legs appear to be like two independent chains this is not the case as they influence each other through slightly complicated Coulomb interactions.

When considering weak interactions a two-leg ladder hopping Hamiltonian must usually be diagonalized so that it can written in terms of two decoupled bands. In our case the two legs are already decoupled in the hopping Hamiltonian making any diagonalization unnecessary. Instead we can immediately make a first approximation by linearizing the lattice fermion operators operators about the Fermi points. This is done by expanding in terms of chiral fields,~\cite{Bunder07}
\begin{align}
d_{q1}(y)/\sqrt{b}\approx&\psi_{Rq}(y)e^{ik_Fy}+
\psi_{L\bar{q}}(y)e^{-ik_Fy}\nonumber\\
d_{q2}(y\pm b-\delta)/\sqrt{b} \approx&\psi_{Rq}(y\pm b)e^{ik_F(y\pm b)}\nonumber\\
&+\psi_{L\bar{q}}(y\pm b)e^{-ik_F(y\pm b)}.
\label{eq:leftright}
\end{align}
On substituting the chiral fields into the Hamiltonian we discard the rapidly varying terms, keeping in mind that the spatial coordinate may now be written as $y=2mb$ for integral $m$. We take the continuum limit, which involves expanding $\psi_{Pq}(y\pm b)$ in a Taylor series about $y$, and retain the lowest order non-zero term. Finally, the Hamiltonian density $\mathcal{H}$ may be obtained by converting the discrete variable $y$ into a continuous variable so that $H_0+H_U+H_V=\int dy[\mathcal{H}_0+\mathcal{H}_U+\mathcal{H}_V]/2b$. The resulting hopping Hamiltonian density is
\begin{equation}
\mathcal{H}_0=v\sum_{q,\alpha}[\psi^{\dag}_{Rq\alpha}i\partial_y \psi_{Rq\alpha}-\psi^{\dag}_{Lq\alpha}i\partial_y \psi_{Lq\alpha}]\label{eq:Hdens0}
\end{equation}
where $v=tb$ is the Fermi velocity. 

The interaction Hamiltonians may be written in terms of the currents 
\begin{align}
J^P_{qq'}&=\sfrac{1}{2}\psi^{\dag}_{Pq\alpha}\psi_{Pq'\alpha},
\qquad &{\bf J}^P_{qq'}&=\sfrac{1}{2}\psi^{\dag}_{Pq\alpha} \bs{\sigma}_{\alpha\beta}\psi_{Pq'\beta}\nonumber\\
I^P_{qq'}&=\sfrac{1}{2}\psi_{Pq\alpha}
\bs{\epsilon}_{\alpha\beta}\psi_{Pq'\beta},
 &{\bf I}^P_{qq'}&=\sfrac{1}{2}\psi_{Pq\alpha} (\bs{\epsilon\sigma})_{\alpha\beta}\psi_{Pq'\beta}.
\label{eq:currents}
\end{align}
After combining both the on-site and nearest neighbour interaction Hamiltonian densities,~\cite{Lin98}
\begin{align}
\mathcal{H}_I=&\mathcal{H}_U+\mathcal{H}_V\nonumber\\
=&b^{\rho}_{qq'}J^R_{qq'}J^L_{qq'}-b^{\sigma}_{qq'} \mathbf{J}^R_{qq'}.\mathbf{J}^L_{qq'}\nonumber\\
&+f^{\rho}_{qq'}J^R_{qq}J^L_{q'q'}
-f^{\sigma}_{qq'}\mathbf{J}^R_{qq}.\mathbf{J}^L_{q'q'}\nonumber\\
&+\textstyle{\frac{1}{2}}[u^{\rho}_{qq'}I^{R\dag}_{qq'} I^L_{\bar{q}\bar{q}'}
-u^{\sigma}_{qq'}\mathbf{I}^{R\dag}_{qq'}. \mathbf{I}^L_{\bar{q}\bar{q}'}+\mathrm{h.c}],\label{eq:HdensI}
\end{align}
where the Hermitian conjugate (h.c.) only refers to the final two terms which are the Umklapp interactions. In deriving this Hamiltonian we took the zeroth order continuum limit of the nearest neighbour interactions. 

The coefficients $f_{qq'}$ and $b_{qq'}$ describe forward and backward scattering, respectively, where $q=\pm=1,2$. To avoid double counting, $f_{qq}=0$. We may also take $u^{\sigma}_{qq}=0$ as $\mathbf{I}_{qq}=0$. Taking Hermiticity and parity symmetry into account gives $b_{12}=b_{21}$ and $f_{12}=f_{21}$ respectively. Therefore we have nine independent coupling strengths which are
\begin{alignat}{2}
b^{\rho}_{11}&=\sfrac{2b}{N_x}(U+3V+3V_{\perp}),\qquad 
&b^{\sigma}_{11}&=\sfrac{2b}{N_x}(U+V-V_{\perp}),\nonumber\\
b^{\rho}_{12}&=\sfrac{2b}{N_x}(U+3V_{\perp}),\qquad
&b^{\sigma}_{12}&=\sfrac{2b}{N_x}(U-2V-V_{\perp}),\nonumber\\
f^{\rho}_{12}&=\sfrac{2b}{N_x}(U+6V+3V_{\perp}),\qquad
&f^{\sigma}_{12}&=\sfrac{2b}{N_x}(U-2V-V_{\perp})\nonumber\\ u^{\rho}_{11}&=\sfrac{2b}{N_x}(U-2V-V_{\perp}),\qquad &&\nonumber\\ 
u^{\rho}_{12}&=\sfrac{2b}{N_x}(2U-V-2V_{\perp}),\qquad
&u^{\sigma}_{12}&=\sfrac{2b}{N_x}3V.
\label{eq:initialcoupling}
\end{alignat}
When written in the form given in Eqs. (\ref{eq:Hdens0},\ref{eq:HdensI}) the armchair and zigzag CNT look identical, but it is important to note that their coupling strengths are quite different and so one would not expect the two nanotubes to have similar solutions.

For further analysis it is convenient to bosonize the Hamiltonian.~\cite{Lin98} In terms of boson fields the fermion operator is
\begin{equation}
\psi_{Pq\alpha}=\kappa_{q\alpha}e^{i\phi_{Pq\alpha}}
\label{eq:bosonization}
\end{equation}
where $P=R/L=\pm$. The boson anticommutation rules are
\begin{align}
[\phi_{Pq\alpha}(y),\phi_{Pq'\beta}(y')]&=iP\pi\delta_{qq'} \delta_{\alpha\beta}\,\mathrm{sgn}(y-y')\nonumber\\
[\phi_{Rq\alpha}(y),\phi_{Lq'\beta}(y')]&=i\pi\delta_{qq'} \delta_{\alpha\beta}
\end{align}
and the Klein factors satisfy
$\{\kappa_{q\alpha},\kappa_{q'\beta}\}=2\delta_{qq'}
\delta_{\alpha\beta}$.
From this we can define a displacement field $\theta_{q\alpha}=\phi_{Rq\alpha}-\phi_{Lq\alpha}$ and a phase field $\varphi_{q\alpha}=\phi_{Rq\alpha}+\phi_{Lq\alpha}$. Then we can define a charge mode $\theta_{q\rho}=(\theta_{q\uparrow}+\theta_{q\downarrow})/\sqrt{2}$ and a spin mode 
$\theta_{q\sigma}=(\theta_{q\uparrow}-\theta_{q\downarrow})/\sqrt{2}$
and likewise for $\varphi$. Finally we define $\theta_{\nu\pm}=(\theta_{1\nu}\pm\theta_{2\nu})/\sqrt{2}$ where $\nu=\rho,\sigma$ and similarly for $\varphi$. 
The Hamiltonian densities in term of the boson fields are
\begin{align}
\mathcal{H}_0=&\frac{v}{8\pi}\sum_{\mu,\pm}[(\partial_y\theta_{\mu\pm})^2 +(\partial_y\varphi_{\mu\pm})^2],\nonumber\\
\mathcal{H}_I=&\frac{1}{32\pi^2}\sum_{\mu,\pm}A_{\mu\pm}[ (\partial_y\theta_{\mu\pm})^2-(\partial_y\varphi_{\mu\pm})^2]
\nonumber\\
&-2\Gamma b_{12}^{\sigma}\cos\varphi_{\rho-}\cos\theta_{\sigma+} \nonumber\\
&+2\cos\theta_{\sigma+}(b^{\sigma}_{11}\cos\theta_{\sigma-} +\Gamma f^{\sigma}_{12}\cos\varphi_{\sigma-})\nonumber\\
&-\cos\varphi_{\rho-}(\Gamma b^+_{12}\cos\theta_{\sigma-} +b^-_{12}\cos\varphi_{\sigma-})\nonumber\\
&-2\Gamma u^{\rho}_{11}\cos\theta_{\rho+}\cos\varphi_{\rho-} -2u^{\sigma}_{12}\cos\theta_{\rho+}\cos\theta_{\sigma+}\nonumber\\ &-\cos\theta_{\rho+}(u^+_{12}\cos\theta_{\sigma-} +\Gamma u^-_{12}\cos\varphi_{\sigma-})]\label{eq:boson-H}
\end{align}
where $A_{\rho\pm}=b^{\rho}_{11}\pm f^{\rho}_{12}$, $A_{\sigma\pm}=-(b^{\sigma}_{11}\pm f^{\sigma}_{12})$, $b^{\pm}_{12}=b^{\sigma}_{12}\pm b^{\rho}_{12}$ and $u^{\pm}_{12}=u^{\sigma}_{12}\pm u^{\rho}_{12}$ and $\Gamma=\kappa_{1\uparrow}\kappa_{1\downarrow} \kappa_{2\uparrow}\kappa_{2\downarrow}$. As $\Gamma^2=1$ we have $\Gamma=\pm 1$, although we shall set $\Gamma=1$.  

\section{Phase analysis}
\label{sec:phases}

\subsection{SO(8) Gross-Neveu model}

We use a well known RG treatment in order to determine the ground state phases of our two-leg ladder.~\cite{Balents96,Konik02,Chen04} The renormalization group flow equations are equivalent to those obtained in Ref. \onlinecite{Lin98} once the slightly different definitions of $J_{Pqq'}$ and $I_{Pqq'}$, as given in Eq. (\ref{eq:currents}), are taken into account. To solve these flow equations we insert the ansatz
\begin{equation}
g^{\nu}_{qq'}(l)=\frac{G^{\nu}_{qq'}}{l_d-l}
\end{equation}
where $l$ is the flow parameter and $g^{\nu}_{qq'}$ is one of the nine independent coupling strengths with the initial values $g^{\nu}_{qq'}(0)$ given in Eq. (\ref{eq:initialcoupling}). The constants $l_d$ and $G^{\nu}_{qq'}$ are obtained on substituting the ansatz into the flow equations. We solve the flow equations numerically 
 for various values of $U$, $V$ and $V_{\perp}$ while always maintaining $|U|>|V|,|V_{\perp}|$. 

\begin{figure}
\begin{center}
\includegraphics[width=8.5cm]{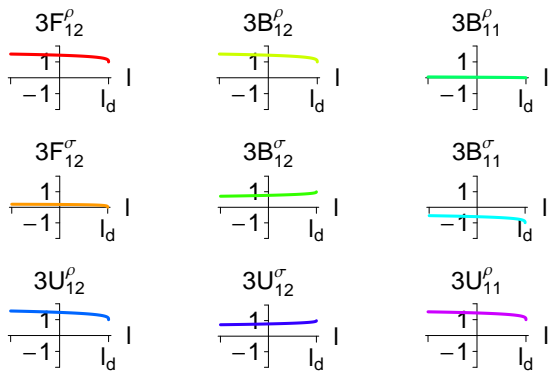}
\caption{Numerical solution of the RG flow equations for the D-Mott phase. The cutoff is $l_d=39.0/U$ and the $l$-axis is over the range $\Delta l=0.2/U$.}
\label{fig:rg1}
\end{center}
\end{figure}

\begin{figure}
\begin{center}
\includegraphics[width=8.5cm]{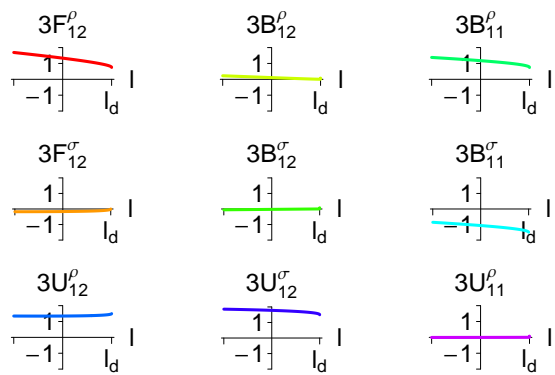}
\caption{Numerical solution of the RG flow equations at the transition between the D-Mott and the S-Mott phases. The cutoff is $l_d=35.2/U$ and the $l$-axis is over the range $\Delta l=4/U$.}
\label{fig:rg2}
\end{center}
\end{figure}

\begin{figure}
\begin{center}
\includegraphics[width=7cm]{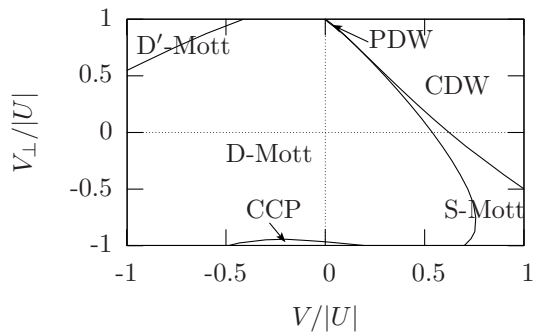}
\caption{Phases for $|V|, |V_{\perp}|<U$, $U>0$.}
\label{fig:phase+}
\end{center}
\end{figure}

\begin{figure}
\begin{center}
\includegraphics[width=7cm]{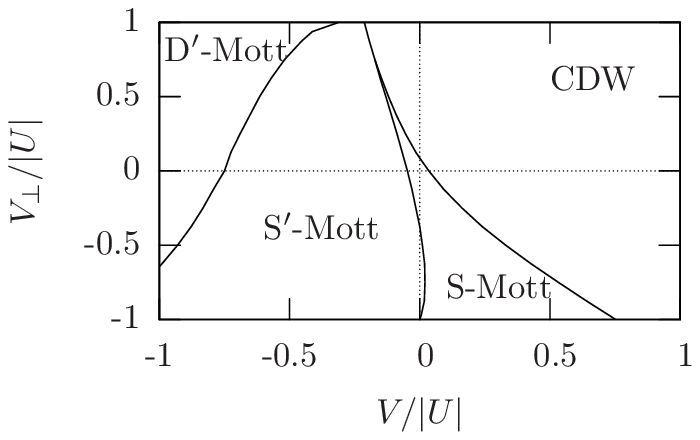}
\caption{Phases for $|V|, |V_{\perp}|<|U|$, $U<0$.}
\label{fig:phase-}
\end{center}
\end{figure}

Two examples of our numerical solutions are shown in Figs. \ref{fig:rg1} and \ref{fig:rg2} in the vicinity of $l=l_d$. What we are really plotting in these nine graphs is $3(l_d-l)g^{\nu}_{qq'}(l)$.  Fig. \ref{fig:rg1} shows that $B_{11}^{\rho}$ and $F_{12}^{\sigma}$ both flow to zero and are therefore negligible. Meanwhile, all other coupling constants flow to the same absolute value. As we shall show later, this solution of the RG flow equations describes the D-Mott phase. Fig. \ref{fig:rg2} describes the phase transition between the D-Mott phase and the S-Mott phase.  In each phase or at each phase transition the nine coupling constants flow to a specific set of values. In total we observe seven distinct phases, as shown in Fig. \ref{fig:phase+} for positive $U$ and Fig. \ref{fig:phase-} for negative $U$. There are only four physically possible phases, the D-Mott, the S-Mott, a CDW and a PDW, i.e, phases obtained for positive (repulsive) interactions, though the PDW exists in a very narrow parameter range. The unphysical phases are the CCP, the D$'$-Mott and the S$'$-Mott. 
The defining coupling constants $G^{\nu}_{qq'}$ of each phase are given in the last column of Table \ref{table:phases}. 
An explanation for the names of the phases will be given in Sec. \ref{name-phases}.

\begin{table*}
\begin{center}
\caption{Each phase's ground state has four pinned fields and four which fluctuate rapidly. For all phases $\langle\theta_{\rho-}\rangle$, $\langle\varphi_{\rho+}\rangle$ and $\langle\varphi_{\sigma+}\rangle$ fluctuate rapidly. The fourth rapidly varying field is indicated in the table by `-'. The phase is determined by the coupling constants. Those coupling constants not mentioned in this table remain small so are negligible.}
\begin{ruledtabular}
\begin{tabular}{c|cccccr@{\extracolsep{0pt}\,$>$\,}l}
phase & $\langle\theta_{\rho+}\rangle$ & $\langle\theta_{\sigma+}\rangle$ & $\langle\theta_{\sigma-}\rangle$ & $\langle\varphi_{\rho-}\rangle$ & $\langle\varphi_{\sigma-}\rangle$ 
& \multicolumn{2}{c}{coupling strength}
\\[3pt]
\hline\\
D-Mott & 0 & 0 & 0 & 0 & - &
$B_{12}^{\rho}=B^{\sigma}_{12}=F^{\rho}_{12}=-B_{11}^{\sigma} =U_{11}^{\rho}=U_{12}^{\rho}=U_{12}^{\sigma}$ & 0\\
S-Mott & 0 & 0 & 0 & $\pi$ & -&
$-B_{12}^{\rho}=-B^{\sigma}_{12}=F^{\rho}_{12}=-B_{11}^{\sigma} =-U_{11}^{\rho}=U_{12}^{\rho}=U_{12}^{\sigma}$ & 0\\
D$'$-Mott & $\pi$ & 0 & 0 & 0 & -&
$B_{12}^{\rho}=B^{\sigma}_{12}=F^{\rho}_{12}=-B_{11}^{\sigma} =-U_{11}^{\rho}=-U_{12}^{\rho}=-U_{12}^{\sigma}$ & 0\\
S$'$-Mott & $\pi$ & 0 & 0 & $\pi$ & -&
$-B_{12}^{\rho}=-B^{\sigma}_{12}=F^{\rho}_{12}=-B_{11}^{\sigma} =U_{11}^{\rho}=-U_{12}^{\rho}=-U_{12}^{\sigma}$ & 0\\
CCP & 0 & 0 & - & 0 & 0&
$-B_{12}^{\rho}=B^{\sigma}_{12}=F^{\rho}_{12}=-F_{12}^{\sigma} =U_{11}^{\rho}=-U_{12}^{\rho}=U_{12}^{\sigma}$ & 0\\
CDW & 0 & 0 & - & $\pi$ & 0&
$B_{12}^{\rho}=-B^{\sigma}_{12}=F^{\rho}_{12}=-F_{12}^{\sigma} =-U_{11}^{\rho}=-U_{12}^{\rho}=U_{12}^{\sigma}$ & 0\\
FDW & $\pi$ & 0 & - & 0 & 0&
$-B_{12}^{\rho}=B^{\sigma}_{12}=F^{\rho}_{12}=-F_{12}^{\sigma} =-U_{11}^{\rho}=U_{12}^{\rho}=-U_{12}^{\sigma}$ & 0\\
PDW & $\pi$ & 0 & - & $\pi$ & 0&
$B_{12}^{\rho}=-B^{\sigma}_{12}=F^{\rho}_{12}=-F_{12}^{\sigma} =U_{11}^{\rho}=U_{12}^{\rho}=-U_{12}^{\sigma}$ & 0\\
\end{tabular}
\end{ruledtabular}
\label{table:phases}
\end{center}
\end{table*}

Most of the zigzag CNT phases (including the unphysical ones) are also known to be armchair CNT phases. The CCP is one phase which has not been explicitly noted in the armchair CNT, but the SF phase in a standard two-leg ladder~\cite{Fjaerestad02,Tsuchiizu02} should map to a CCP in the armchair CNT. One phase which can in principle exist in a zigzag CNT but which we do not find in our phase diagrams, Figs. \ref{fig:phase+} and \ref{fig:phase-}, is the FDW. Although, we do not observe this phase for the half-filling case considered here, we will later show that it can be found in phase diagrams of doped zigzag CNT.~\cite{Tsuchiizu02,Tsuchiizu05}

If the coupling strengths for the D-Mott phase are substituted into Eq. (\ref{eq:boson-H}) it can be shown that, if we define
\begin{alignat}{2}
(\theta,\varphi)_1&=(\theta,\varphi)_{\rho+},\qquad
(\theta,\varphi)_2&=(\theta,\varphi)_{\sigma+}\nonumber\\
(\theta,\varphi)_3&=(\theta,\varphi)_{\sigma-},\qquad
(\theta,\varphi)_4&=(\varphi,\theta)_{\rho-}\label{eq:DMott}
\end{alignat}
then
\begin{align}
\mathcal{H}_0=&\frac{v}{8\pi}\sum_a[(\partial_y\theta_a)^2 +(\partial_y\varphi_a)^2]\nonumber\\
\mathcal{H}_I=&-\frac{g}{2\pi^2}\sum_a\partial_y\phi_{Ra} \partial_y\phi_{La}-4\pi g\sum_{a\neq b}\cos\theta_a\cos\theta_b
\label{eq:DMottH}
\end{align}
where we have defined a chiral field $\phi_{Pa}=(\varphi_a+P\theta_a)/2$ and $4g=|G^{\nu}_{qq'}|$ for $G^{\nu}_{qq'}\neq 0$. In the semiclassical limit the ground state can be determined by minimizing the above Hamiltonian. It is not hard to see that the Hamiltonian will be minimized when we simultaneously pin  either $\theta_a=2n_a\pi$ or $\theta_a=(2n_a+1)\pi$ for all $a$ and integral $n_a$. These pinned fields describe the gapped (or massive) excitations of the system as any change in $\theta_a$ must be finite. The system is unaffected by any change in the unpinned fields, even if the changes are infinitesimally small, so these fields define the gapless (or massless) excitations. 
The $\varphi_a$ field is dual to $\theta_a$, which means if one field is pinned the other must vary rapidly in the semiclassical limit i.e., the Heisenberg uncertainty principle comes into effect and causes one field to vary rapidly if its dual field is well known.

We now show how the bosonized Hamiltonian may be mapped onto an SO(8) GN model.~\cite{Lin98} Firstly we refermionize using
\begin{align}
\psi_{Pa}&=\kappa_a e^{i\phi_{Pa}},\qquad a=1,2,3\nonumber\\
\psi_{P4}&=P\kappa_4 e^{i\phi_{P4}}
\end{align}
where the Klein factors are
$\kappa_1=\kappa_{2\uparrow}$, $\kappa_2=\kappa_{1\uparrow}$,
$\kappa_3=\kappa_{1\downarrow}$, and $\kappa_4=\kappa_{2\downarrow}$.
We then map to Majorana fields,
\begin{equation}
\psi_{Pa}=(\eta_{P2a}+i\eta_{P2a-1})/\sqrt{2}
\end{equation}
to obtain the SO(8) GN model,
\begin{equation}
\mathcal{H}=\half\eta_{RA}i\partial_y\eta_{RA}
-\half\eta_{LA}i\partial_y\eta_{LA}
+g G_R^{AB}G_L^{AB}
\end{equation}
with $A,B=1,2\ldots, 8$ and currents
$G_P^{AB}=\eta_{PA}\eta_{PB}$ for $A\neq B$,
thus showing that the D-Mott phase of the two-leg ladder has its symmetry enlarged to SO(8).  

The Hamiltonians of the seven remaining phases in terms of $\theta_a$ and $\varphi_a$ or $\eta_{Pa}$ may be obtained using appropriate mappings from the D-Mott phase, thereby showing that all phases have an SO(8) symmetry, although they do not share the same SO(8) symmetry. If $\theta_a$ and $\varphi_a$ are as defined for the D-Mott phase in Eq. (\ref{eq:DMott}) then for the D$'$-Mott phase,
\begin{equation}
\theta^{D'}_a=\left\{\begin{array}{ll}
\theta_a+\pi,\qquad & a=1\\
\theta_a, & a=2,3,4
\end{array}\right.\label{eq:D'Mottmap}
\end{equation}
and $\varphi^{D'}_a=\varphi_a$. In terms of the GN Majorana fields,
\begin{equation}
\eta^{D'}_{PA}=\left\{\begin{array}{ll}
P\eta_{PA},\qquad & A=1,2\\
\eta_{PA},& A=3,\ldots,8.
\end{array}\right.
\end{equation}
Therefore the two D-Mott phases share an SO(6) subalgebra. 
Similarly for the S-Mott,
\begin{equation}
\theta^{S}_a=\left\{\begin{array}{ll}
\theta_a,& a=1,2,3\\
\theta_a+\pi,\qquad & a=4
\end{array}\right.
\end{equation}
and $\varphi^{S}_a=\varphi_a$. In terms of the GN Majorana fields,
\begin{equation}
\eta^{S}_{PA}=\left\{\begin{array}{ll}
\eta_{PA},\qquad & A=1,2,3,4,5,6\\
P\eta_{PA},& A=7,8.
\end{array}\right.
\end{equation}
Therefore the D-Mott and the S-Mott also share an SO(6) subalgebra, although it is different from the subalgebra shared by the D-Mott and the D$'$-Mott. All the symmetries between all the phases are shown in Table \ref{table:symmetries}.

\begin{table*}
\begin{center}
\caption{Symmetries shared by different phases. The upper right corner is for the undoped case and the lower left corner is for the doped case.}
\begin{ruledtabular}
\begin{tabular}{c|cccccccc}
phase & D-Mott/SC & D$'$-Mott & S-Mott/SC & S$'$-Mott & CCP & FDW & CDW & PDW
\\[3pt]
\hline\\
D-Mott/SC & & SO(6) & SO(6) & SO(4) & SO(7) & SO(5) & SO(5) & SO(3)\\
D$'$-Mott & &  & SO(4) & SO(6) & SO(5) & SO(7) & SO(3) & SO(5)\\
S-Mott/SC & SO(4) &  &  & SO(6) & SO(5) & SO(3) & SO(7) & SO(5)\\
S$'$-Mott & &  &  & & SO(3) & SO(5) & SO(5) & SO(7)\\
CCP  & SO(5) &  & SO(3) & &  & SO(6) & SO(6) & SO(4)\\
FDW  & &  &  & &  &  & SO(4) & SO(6)\\
CDW  & SO(3) &  & SO(5) & & SO(4)  & &  & SO(6)
\end{tabular}
\end{ruledtabular}
\label{table:symmetries}
\end{center}
\end{table*}

\subsection{Classification of phases}
\label{name-phases}
In Sec. \ref{intro} we gave a qualitative description of the eight phases, while in the previous section the phases were defined in terms of their pinned fields. In this section we relate the pinned fields to the qualitative description. We do this by calculating various order parameters such as the current, kinetic energy, electron density, superconducting order parameter and the pair field operator. 

\subsubsection{Electron density}
Each lattice site contributes one electron so the average electron density per lattice site is one. Deviations from this average value can be define by
\begin{equation}
n(\mathbf{R'})=\sum_{\alpha}c^{\dag}_{i\alpha}(\mathbf{R'})
c_{i\alpha}(\mathbf{R'})
\end{equation}
on a generic site of the carbon lattice with $i=1,2$.
After using the same mappings which provided us with the two-leg ladder Hamiltonian and then bosonizing the resulting equation we can represent the deviation from average density by
\begin{align}
n(m)=&(-1)^m\sum_{\alpha,q}\kappa_{q\alpha}\kappa_{\bar{q}\alpha} [e^{-i\phi_{Rq\alpha}+i\phi_{L\bar{q}\alpha}}\nonumber\\
&-e^{-i\phi_{Lq\alpha}+i\phi_{R\bar{q}\alpha}}]\nonumber\\
=&-i8(-1)^{m}\kappa_{1\uparrow}\kappa_{2\uparrow}\nonumber\\
&\times(\sin\half\varphi_{\rho-}\cos\half\varphi_{\sigma-}
\cos\half\theta_{\rho+}\cos\half\theta_{\sigma+}\nonumber\\
&-\cos\half\varphi_{\rho-}\sin\half\varphi_{\sigma-}
\sin\half\theta_{\rho+}\sin\half\theta_{\sigma+}).
\end{align}
The Klein factors have been simplified by using $(\kappa_{2\uparrow}\kappa_{1\uparrow})
(\kappa_{1\uparrow}\kappa_{2\uparrow}) =1$ and $\Gamma=1$.
To evaluate the electron density we consult Table \ref{table:phases} and substitute in the ground state values. For all Mott phases $\varphi_{\rho-}$ is rapidly varying so $n(m)=0$ and there is no deviation from the average electron density of one electron per site.  
The deviation from the average electron density is also zero for all spin wave phases except the CDW. In the CDW $n(m)\propto(-1)^m$. Therefore, as shown in Fig. \ref{fig:phases}(e) the CDW consists of pairs of electrons on alternate sites. Note that there are two possible forms of the CDW, one where all electrons are positioned on the sublattice represented by green triangles, and the other where all electrons are positioned on the sublattice represented by red squares.

\subsubsection{Current between nearest neighbours}
The current between nearest neighbours on the carbon lattice may be defined as
\begin{align}
j_{\perp}(\mathbf{R})&=-i\sum_{\alpha}[c^{\dag}_{1\alpha}(\mathbf{R}) c_{2\alpha}(\mathbf{R}+\mathbf{d})-h.c]
\nonumber\\
j_{1}(\mathbf{R})&=i\sum_{\alpha}[c^{\dag}_{1\alpha}(\mathbf{R}) c_{2\alpha}(\mathbf{R}+\mathbf{d}_+)-h.c]
\nonumber\\
j_{2}(\mathbf{R})&=i\sum_{\alpha}[c^{\dag}_{1\alpha}(\mathbf{R})
c_{2\alpha}(\mathbf{R}+\mathbf{d}_-)-h.c].
\label{eq:current}
\end{align}
We have defined the current to always have a negative $y$ component so $j_{\perp}(\mathbf{R})$ is travelling away from $\mathbf{R}$ along the perpendicular while $j_{1,2}(\mathbf{R})$ are travelling towards $\mathbf{R}$ along the zigzags. 
Note that when calculating order parameters which act between different sites we use ${\bf R}$ which describes all lattice sites in one sublattice of the CNT. In contrast, when calculating the order parameters which act on one site, such as the electron density, we use ${\bf R}'$ which describes all lattice sites in both sublattices.  
After the usual mappings we find $j_{\perp}=0$ and, since $R_y=2mb$, 
\begin{align}
j_1(2m)=&-j_2(2m)=i\sqrt{3}\sum_{\alpha,q}\,\mathrm{sgn}(q)
\kappa_{q\alpha}\kappa_{\bar{q}\alpha}\nonumber\\
&\times[
e^{-i\phi_{Rq\alpha}+i\phi_{L\bar{q}\alpha}}
+e^{i\phi_{Rq\alpha}-i\phi_{L\bar{q}\alpha}}]\nonumber\\
=&i8\sqrt{3}\kappa_{1\uparrow}\kappa_{2\uparrow}\nonumber\\
&\times(\cos\half\varphi_{\rho-}\cos\half\varphi_{\sigma-}
\cos\half\theta_{\rho+}\cos\half\theta_{\sigma+}\nonumber\\
&+\sin\half\varphi_{\rho-}\sin\half\varphi_{\sigma-}
\sin\half\theta_{\rho+}\sin\half\theta_{\sigma+}).
\end{align}
This current is a function of $2m$ rather than $m$ because we have defined the current in terms of one sublattice rather than the entire lattice.  

Table \ref{table:phases} shows that
for all Mott phases $\varphi_{\rho-}$ is rapidly varying so $j_{1,2}=0$. In all other phases, except the CCP we also find $j_{1,2}=0$. In the CCP $j_1=-j_2=i8\sqrt{3}\kappa_{1\uparrow}\kappa_{2\uparrow}$. This describes currents running along the zigzags of the CNT. Note that although the carbon lattice has a non-zero flow of current along the zigzags, all currents in the equivalent two-leg ladder vanish. This is because the total current in the $y$ direction is zero in the CNT and the legs of the two-leg ladder are in the $y$ direction. We have shown that there are two possibilities for the CDW (positive electron densities on one of two sublattices) and similarly, there are two possibilities for the CCP with currents flowing in one of two directions about the CNT. Note that currents along different zigzags is always in the same direction, i.e., either clockwise or anti-clockwise and not both in the same CNT.

If we constructed currents similar to Eq. (\ref{eq:current}) for an armchair CNT we would find a different type of CCP where currents still flow around the tube, but along the `armchair' bonds. In the equivalent two-leg ladder we would find a SF phase. The equivalent two-leg ladder phase of the zigzag CNT's CCP cannot strictly speaking be classified as a SF phase as there are no true plaquettes around which current can flow. However, we could possibly define a plaquette in this two-leg ladder as being between two nearest neighbours with an infinitesimally narrow width in the $x$ direction. Then current flow around a plaquette is equivalent to equal currents flowing back and forth between  two sites and cancelling each other out, as shown in Fig. \ref{fig:phases}(g).

\subsubsection{Kinetic energy between nearest neighbours}
The kinetic energy is defined similarly to the current
\begin{align}
B_{\perp}(\mathbf{R})&=i\sum_{\alpha}[c^{\dag}_{1\alpha}(\mathbf{R})
c_{2\alpha}(\mathbf{R}+\mathbf{d})+h.c]
\nonumber\\
B_{1}(\mathbf{R})&=i\sum_{\alpha}[c^{\dag}_{1\alpha}(\mathbf{R})
c_{2\alpha}(\mathbf{R}+\mathbf{d}_+)+h.c]
\nonumber\\
B_{2}(\mathbf{R})&=i\sum_{\alpha}[c^{\dag}_{1\alpha}(\mathbf{R})
c_{2\alpha}(\mathbf{R}+\mathbf{d}_-)+h.c].
\end{align}
After the appropriate mappings we find $B_{1}(2m)=B_2(2m)=-B_{\perp}(2m)/2$ and
\begin{align}
B_{1}(2m)=&\sum_{q,\alpha}\kappa_{q\alpha}\kappa_{\bar{q}\alpha} (e^{-i\phi_{Rq\alpha}+i\phi_{L\bar{q}\alpha}}+
e^{i\phi_{Rq\alpha}-i\phi_{L\bar{q}\alpha}})\nonumber\\
=&-8\kappa_{1\uparrow}\kappa_{2\uparrow}\nonumber\\
&\times(\sin\half\theta_{\rho+}\cos\half\theta_{\sigma+} \sin\half\varphi_{\rho-}\cos\half\varphi_{\sigma-}\nonumber\\
&+\cos\half\theta_{\rho+}\sin\half\theta_{\sigma+} \cos\half\varphi_{\rho-}\sin\half\varphi_{\sigma-})
\end{align}
which vanishes in all phases except the PDW. In the PDW $B_1(2m)=-8\kappa_{1\uparrow}\kappa_{2\uparrow}$. Like the CDW and the CCP the PDW also has two possibilities. One possibility has positive kinetic energy on the diagonal bonds but negative on the vertical bonds, and the other possibility has the signs exchanged.   

\subsubsection{Current between next-nearest neighbours}
We can define six currents between next-nearest neighbours,
\begin{align}
j_{11}(\mathbf{R})=& i\sum_{\alpha}[c^{\dag}_{1\alpha}
(\mathbf{R}+\mathbf{a}_-)c_{1\alpha}(\mathbf{R})-h.c]
\nonumber\\
j_{12}(\mathbf{R})=&i\sum_{\alpha}[c^{\dag}_{1\alpha}(\mathbf{R})
c_{1\alpha}(\mathbf{R}+\mathbf{a}_+)-h.c]
\nonumber\\
j_{1x}(\mathbf{R})=&i\sum_{\alpha}[c^{\dag}_{1\alpha}(\mathbf{R}+
\mathbf{a}_-)c_{1\alpha}(\mathbf{R}+\mathbf{a}_+)-h.c]
\nonumber\\
j_{21}(\mathbf{R})=&i\sum_{\alpha}[c^{\dag}_{2\alpha}
(\mathbf{R}+\mathbf{d}_-)
c_{2\alpha}(\mathbf{R}+\mathbf{d})-h.c]
\nonumber\\
j_{22}(\mathbf{R})=&i\sum_{\alpha}[c^{\dag}_{2\alpha}
(\mathbf{R}+\mathbf{d})
c_{2\alpha}(\mathbf{R}+\mathbf{d}_+)-h.c]
\nonumber\\
j_{2x}(\mathbf{R})=&i\sum_{\alpha}[c^{\dag}_{2\alpha}
(\mathbf{R}+\mathbf{d}_-)
c_{2\alpha}(\mathbf{R}+\mathbf{d}_+)-h.c].
\end{align}
The first three currents act between sites of the first sublattice and the other currents act between sites of the second sublattice. Currents $j_{1x}$ and $j_{2x}$ are in the $x$ direction. All currents are defined to have negative $x$ components. 
On expanding in terms of $\psi_{R/L}$ we can see that $j_{11}=j_{12}=-j_{1x}=j_{21}=j_{22}=-j_{2x}$.
Bosonizing gives
\begin{align}
j_{11}(2m)=&\sqrt{3}\sum_{q\alpha}\,\mathrm{sgn}(q)
\kappa_{q\alpha}\kappa_{\bar{q}\alpha}\nonumber\\
&\times[
e^{-i\phi_{Rq\alpha}+i\phi_{Lq\alpha}}
-e^{i\phi_{Rq\alpha}-i\phi_{Lq\alpha}}]\nonumber\\
=&-i8\sqrt{3}\kappa_{1\uparrow}\kappa_{2\uparrow}
\nonumber\\
&\times
(\cos\half\varphi_{\rho-}\cos\half\varphi_{\sigma-}
\sin\half\theta_{\rho+}\cos\half\theta_{\sigma+}
\nonumber\\
&-\sin\half\varphi_{\rho-}\sin\half\varphi_{\sigma-}
\cos\half\theta_{\rho+}\sin\half\theta_{\sigma+}).
\end{align}
These currents vanish in all but the FDW. In the FDW $j_{11}(2m)=-i8\sqrt{3}\kappa_{1\uparrow}\kappa_{2\uparrow}$ which is equal in magnitude to the nearest neighbour current $j_1$ in the CCP.  Like the CCP all currents in the two-leg ladder vanish. 

Like the other charge density waves the FDW has two possibilities. The current about one sublattice can either be clockwise or anti-clockwise and the current about the other sublattice must then be either anticlockwise or clockwise, respectively. One could define this phase to be a type of CNT SF phase where the plaquettes are the triangles of either sublattice in the CNT. Both sublattices carry their own staggered current flux.~\cite{Nersesyan03}

\subsubsection{Kinetic energy between next-nearest neighbours}
 The kinetic energy between next-nearest neighbours is similar to the current between next-nearest neighbours. We define
\begin{align}
B_{11}(\mathbf{R})=& i\sum_{\alpha}[c^{\dag}_{1\alpha}
(\mathbf{R}+\mathbf{a}_-)c_{1\alpha}(\mathbf{R})+h.c]
\nonumber\\
B_{12}(\mathbf{R})=&i\sum_{\alpha}[c^{\dag}_{1\alpha}(\mathbf{R})
c_{1\alpha}(\mathbf{R}+\mathbf{a}_+)+h.c]
\nonumber\\
B_{1x}(\mathbf{R})=&i\sum_{\alpha}[c^{\dag}_{1\alpha}(\mathbf{R}+
\mathbf{a}_-)c_{1\alpha}(\mathbf{R}+\mathbf{a}_+)+h.c]
\nonumber\\
B_{21}(\mathbf{R})=&i\sum_{\alpha}[c^{\dag}_{2\alpha}
(\mathbf{R}+\mathbf{d}_-)
c_{2\alpha}(\mathbf{R}+\mathbf{d})+h.c]
\nonumber\\
B_{22}(\mathbf{R})=&i\sum_{\alpha}[c^{\dag}_{2\alpha}
(\mathbf{R}+\mathbf{d})
c_{2\alpha}(\mathbf{R}+\mathbf{d}_+)+h.c]
\nonumber\\
B_{2x}(\mathbf{R})=&i\sum_{\alpha}[c^{\dag}_{2\alpha}
(\mathbf{R}+\mathbf{d}_-)
c_{2\alpha}(\mathbf{R}+\mathbf{d}_+)+h.c].
\end{align}
We can show that all the kinetic energies are equal,
$B_{11}=B_{12}=B_{1x}=B_{21}=B_{22}=B_{2x}$ and
\begin{align}
B_{11}(2m)=&-i\sum_{q\alpha}[
\kappa_{q\alpha}\kappa_{\bar{q}\alpha}
e^{-i\phi_{Rq\alpha}+i\phi_{Lq\alpha}}\nonumber\\
&-\kappa_{q\alpha}\kappa_{\bar{q}\alpha}
e^{i\phi_{Rq\alpha}-i\phi_{Lq\alpha}}]\nonumber\\
=&-8\kappa_{1\uparrow}\kappa_{2\uparrow}\nonumber\\
&\times(\sin\half\varphi_{\rho-}\cos\half\varphi_{\sigma-}
\cos\half\theta_{\rho+}\cos\half\theta_{\sigma+}\nonumber\\
&-\cos\half\varphi_{\rho-}\sin\half\varphi_{\sigma-}
\sin\half\theta_{\rho+}\sin\half\theta_{\sigma+})
\end{align}
which vanishes for all phases except the CDW where $B_{11}(2m)=-8\kappa_{1\uparrow}\kappa_{2\uparrow}$, which is equal to the kinetic energy $B_{1,2}$ in the PDW. We have not included this kinetic energy in the CDW illustration given in Fig. \ref{fig:phases}(e) to avoid cluttering the picture. This kinetic energy in the CDW joins sites with equal electron density. 

\subsubsection{Superconducting order parameter and pair field operator}

The S-SC order parameter describes pairings on the same site of the CNT lattice, 
\begin{equation}
\Delta_{S}(\mathbf{R}')=c_{i\uparrow}(\mathbf{R}')
c_{i\downarrow}(\mathbf{R}')=
\sum_q[\Delta_{Rq}+\Delta_{L\bar{q}}]\label{eq:scopS}
\end{equation}
where
$\Delta_{Pq}=\psi_{Pq\uparrow}\psi_{\bar{P}q\downarrow}$ is the pair field operator in leg $q$ of the two-leg ladder.
The D-SC order parameter describes pairings across neighbouring sites and can also be written in terms of the pair field operator,
\begin{align}
\Delta_{D\perp}(\mathbf{R})=&c_{1\uparrow}(\mathbf{R})
c_{2\downarrow}(\mathbf{R}+\mathbf{d})
=i\sum_q[\Delta_{Rq}-\Delta_{L\bar{q}}]\nonumber\\
\Delta_{D1}(\mathbf{R})=&c_{1\uparrow}(\mathbf{R})
c_{2\downarrow}(\mathbf{R}+\mathbf{d}_+)\nonumber\\
=&-i\sum_q[\Delta_{Rq}-\Delta_{L\bar{q}}]
e^{-iq\pi/3}\nonumber\\
\Delta_{D2}(\mathbf{R})=&c_{1\uparrow}(\mathbf{R})
c_{2\downarrow}(\mathbf{R}+\mathbf{d}_-)
\nonumber\\
=&-i\sum_q[\Delta_{Rq}-\Delta_{L\bar{q}}]
e^{iq\pi/3}.\label{eq:scopD}
\end{align}
Recall that $\mathbf{R}$ describes only one sublattice, while $\mathbf{R}'$ describes the entire CNT lattice.

The singlet pairing between electrons in the two-leg ladder is defined by the sign of $\langle\Delta_{P1}\Delta_{P2}^{\dag}\rangle$. A negative value indicates a $d$-wave symmetry while a positive value indicates a $s$-wave symmetry. We define this symmetry in terms of the equivalent two-leg ladder, in analogy with the standard two-leg ladder.~\cite{Lin98} 
For all four Mott phases the fields $\theta_{\sigma\pm}$ are pinned which implies that the two spin fields $\theta_{q\sigma}$ are also pinned. Having $\theta_{q\sigma}$ pinned indicates that excitations in the $q$th leg of the two-leg ladder with non-zero spin require energy, implying singlet pairings in the Mott phases. For the D-Mott and the D$'$-Mott phases $\langle\Delta_{P1}\Delta_{P2}^{\dag}\rangle<0$ so they have $d$-wave symmetry. For the two S-Mott phases however $\langle\Delta_{P1}\Delta_{P2}^{\dag}\rangle>0$ so they have $s$-wave symmetry. This explains the prefixes of the four Mott phases. 
In the four density waves $\varphi_{\sigma-}$ is pinned rather than $\theta_{\sigma-}$  so we can make no conclusion about the total spin in each leg. Because $\theta_{\sigma-}$ is dual to $\varphi_{\sigma-}$ it varies rapidly in the density waves and  $\langle\Delta_{P1}\Delta_{P2}^{\dag}\rangle=0$.

From Table \ref{table:phases} it can be seen that the only difference between the D-Mott (S-Mott) and the D$'$-Mott (S$'$-Mott) is the change in $\theta_{\rho+}$ from $0$ to $\pi$. The field $\theta_{\rho+}$ describes the charge gap and indicates that the singlet's center of mass is shifted by $\pi$, or half a unit cell, when comparing the D-Mott (S-Mott) to the D$'$-Mott (S$'$-Mott). In the S-Mott the electron pairs oscillate between nearest neighbours. An S$'$-Mott is obtained from the S-Mott by shifting the centre of mass of each electron pair by half a unit cell so the S$'$-Mott phase must describe pairings which oscillate between next-nearest neighbours. The relationship between the D-Mott and the D$'$-Mott is similar.

We now return to the superconducting order parameters in Eqs. (\ref{eq:scopS}) and (\ref{eq:scopD}). On bosonizing it can be seen that the S-SC order parameter is
\begin{align}
\Delta_S(m)=&\sum_q\kappa_{q\uparrow}\kappa_{q\downarrow}[
e^{i\phi_{Rq\uparrow}+i\phi_{Lq\downarrow}}
+e^{i\phi_{Rq\downarrow}+i\phi_{Lq\uparrow}}]\nonumber\\
=&4\kappa_{1\uparrow}\kappa_{1\downarrow}e^{i\varphi_{\rho+}/2}
\nonumber\\
&\times(-\cos\half\varphi_{\rho-} \sin\half\theta_{\sigma+}\sin\half\theta_{\sigma-}\nonumber\\
&+i\sin\half\varphi_{\rho-}\cos\half\theta_{\sigma+} \cos\half\theta_{\sigma-})].
\end{align}
The D-SC order parameters are
\begin{align}
\Delta_{D\perp}(2m)=&i\sum_q[
\kappa_{q\uparrow}\kappa_{q\downarrow}
e^{i\phi_{Rq\uparrow}+i\phi_{Lq\downarrow}}\nonumber\\
&-\kappa_{\bar{q}\uparrow}\kappa_{\bar{q}\downarrow}
e^{i\phi_{R\bar{q}\downarrow}+i\phi_{L\bar{q}\uparrow}}]
\nonumber\\
=&-4\kappa_{1\uparrow}\kappa_{1\downarrow}e^{i\varphi_{\rho+}/2}
\nonumber\\
&\times(\cos\half\varphi_{\rho-}\cos\half\theta_{\sigma+} \sin\half\theta_{\sigma-}\nonumber\\
&+i\sin\half\varphi_{\rho-} \sin\half\theta_{\sigma+}\cos\half\theta_{\sigma-})
\end{align}
and if $j=1,2$
\begin{align}
\Delta_{Dj}(2m)=&-i\sum_q[
\kappa_{q\uparrow}\kappa_{q\downarrow}
e^{i\phi_{Rq\uparrow}+i\phi_{Lq\downarrow}}\nonumber\\
&-\kappa_{\bar{q}\uparrow}\kappa_{\bar{q}\downarrow}
e^{i\phi_{R\bar{q}\downarrow}+i\phi_{L\bar{q}\uparrow}}]
e^{(-1)^jiq\pi/3}\nonumber\\
=&2\kappa_{1\uparrow}\kappa_{1\downarrow}e^{i\varphi_{\rho+}/2}
\nonumber\\
&\times(\cos\half\varphi_{\rho-}\cos\half\theta_{\sigma+} 
\sin\half\theta_{\sigma-}\nonumber\\
&+i\sin\half\varphi_{\rho-}\sin\half\theta_{\sigma+} 
\cos\half\theta_{\sigma-}\nonumber\\
&+(-1)^j\sqrt{3}(\cos\half\varphi_{\rho-}
\cos\half\theta_{\sigma+} 
\cos\half\theta_{\sigma-}\nonumber\\
&-i\sin\half\varphi_{\rho-}\sin\half\theta_{\sigma+} 
\sin\half\theta_{\sigma-})).
\end{align}
Because of the presence of the rapidly varying $\varphi_{\rho+}$ all superconducting order parameters will vanish in all phases.

\subsection{Phase transitions}

When numerically solving the RG flow equations we find seven phases defined by the coupling strengths $G^{\nu}_{qq'}$. The transitions between these phases are also defined by a unique set of coupling strengths as given in Table \ref{table:transitions}. From this table it can be seen that although we observed eight different transitions, there were only three types of transitions, defined as Gaussian, Ising and SO(5) GN$\times$SO(3) WZW. The Gaussian and Ising transitions were discussed in Ref. \onlinecite{Lin98}. 

\begin{table*}
\begin{center}
\caption{The phase transitions described here refer only to the transitions displayed in Figs. \ref{fig:phase+} and \ref{fig:phase-} and are not exhaustive. For example, one would expect a Gaussian transition between a CCP and a FDW and an Ising transition between a PDW and an S$'$-Mott phase, but as we did not find these transitions in our phase diagram we will not discuss them further.}
\begin{ruledtabular}
\begin{tabular}{ccc}
phase transition & coupling strength & classification
\\[3pt]
\hline\\
D-Mott $\leftrightarrow$ S-Mott & $F_{12}^{\rho}=B^{\rho}_{11}=1/4,$ $U^{\rho}_{12}=U^{\sigma}_{12}=-B^{\sigma}_{11}=1/2$ &
Gaussian\\
D-Mott $\leftrightarrow$ D$'$-Mott &
$F_{12}^{\rho}=B^{\rho}_{11}=1/4$, $B_{12}^{\rho}=B^{\sigma}_{12}=-B^{\sigma}_{11}=1/2$ &
Gaussian\\
S-Mott $\leftrightarrow$ S$'$-Mott &
$-F_{12}^{\rho}=B^{\rho}_{11}=1/4$, $B_{12}^{\rho}=B^{\sigma}_{12}=B^{\sigma}_{11}=-1/2$ &
Gaussian\\
D$'$-Mott $\leftrightarrow$ S$'$-Mott & $F_{12}^{\rho}=B^{\rho}_{11}=1/4,$ $U^{\rho}_{12}=U^{\sigma}_{12}=B^{\sigma}_{11}=-1/2$ &
Gaussian\\
PDW $\leftrightarrow$ CDW & $F_{12}^{\rho}=-B^{\rho}_{11}=1/4,$ $B^{\rho}_{12}=-F^{\sigma}_{12}=-B^{\sigma}_{12}=1/2$ &
Gaussian\\
D-Mott $\leftrightarrow$ CCP &
$F^{\rho}_{12}=B^{\sigma}_{12}=U^{\sigma}_{12}=U^{\rho}_{11}
=2/5$, $F^{\sigma}_{12}=B^{\sigma}_{11}=-1/5$ & Ising \\
S-Mott $\leftrightarrow$ CDW &
$F^{\rho}_{12}=-B^{\sigma}_{12}=U^{\sigma}_{12}=-U^{\rho}_{11}
=2/5$, $F^{\sigma}_{12}=B^{\sigma}_{11}=-1/5$ & Ising \\
D-Mott $\leftrightarrow$ PDW &
$F^{\rho}_{12}=B^{\rho}_{12}=U^{\rho}_{12}=U^{\rho}_{11}=2/3$ &
SO(5) GN$\times$SO(3) WZW
\end{tabular}
\end{ruledtabular}
\label{table:transitions}
\end{center}
\end{table*}

We can make several generalizations about what type of phase transition should exist between two particular phases.
If two phases have the same pinned fields in the ground state and all but one of these fields are pinned to the same value then the transition between these two phases is Gaussian. For example, in the D-Mott phase $\theta_{\rho+}=0$ and in the D$'$-Mott phase $\theta_{\rho+}=\pi$, but all other pinned fields have the same values. However, if the coupling constants of the D-Mott $\leftrightarrow$ D$'$-Mott transition shown in Table \ref{table:transitions} are substituted into Eq. (\ref{eq:boson-H}) no sinusoidal term involving either $\theta_{\rho+}$ or its dual field $\varphi_{\rho+}$ appears and therefore these fields are massless at the transition. Those fields which are pinned to the same value in both the D-Mott and the D$'$-Mott phases remain in the Hamiltonian so are massive. 
Refermionizing this new Hamiltonian and mapping to Majorana fermions gives an SO(6) GN model. In other words, this phase transition is described by a single gapless bosonic mode, which implies a central charge $c=1$. Such a phase transition is Gaussian. By comparing changes in pinned fields is not hard to see why the first six transitions in Table \ref{table:transitions} must all be of the same type.
%To analyse the transition we construct an effective Hamiltonian in terms of the massless field $\theta_{\rho+}$ by integrating out all the massive fields.~\cite{Lin98} This effective semiclassical low energy Hamiltonian has a term of the form $-\lambda\cos\theta_1$, where $\theta_1=\theta_{\rho+}$. If $\theta_1=0$ the ground state requires $\lambda>0$ but if $\theta_1=\pi$ the ground state requires $\lambda<0$. Therefore the phase transition is described by a single gapless bosonic mode, which implies a central charge $c=1$. A phase transition defined in terms of a continuous variable $\lambda$ and a central charge $c=1$ is Gaussian.  

When two dual fields exchange their pinned value we have an Ising transition. For example, the D-Mott phase and the CCP have the same pinned values except in the D-Mott phase $\theta_{\sigma-}=0$ and $\varphi_{\sigma-}$ is rapidly varying and in the CCP  $\varphi_{\sigma-}=0$ and $\theta_{\sigma-}$ is rapidly varying. If we substitute the coupling constants from Table \ref{table:transitions} into Eq. (\ref{eq:boson-H}) we obtain a rather complicated Hamiltonian. We can integrate out all modes which are massive at the critical point (i.e, those terms only containing fields which do not change their values across the transition) to obtain an effective interaction Hamiltonian.~\cite{Lin98} This effective Hamiltonian is refermionized and mapped to Majorana fermions and we obtain a GN model with one gapped Majorana fermion and one gapless. The discarded massive modes provide six more gapped modes. Therefore, these types of transitions have a single gapless fermion indicating central charge of $c=1/2$, which defines an Ising transition.

We have one remaining phase transition, between the D-Mott phase and the PDW. This case initially appears to be more complicated than the Ising or Gaussian transitions as we now have four fields which change their pinned value during the transition,  $\theta_{\rho+}$, $\varphi_{\rho-}$, $\theta_{\sigma-}$ and $\varphi_{\sigma-}$. Again we substitute the relevant coupling constants from Table \ref{table:transitions} into Eq. (\ref{eq:boson-H}). As several coupling constants are zero we obtain a relatively simple Hamiltonian and there is no real need to integrate out the massive modes, as was done for the Ising transition. One can proceed as one did for the Gaussian transition and immediately refermionize and then map to Majorana fermions. The resulting interaction Hamiltonian is of the form
\begin{equation}
\mathcal{H}_{I\mathrm{D-Mott}\leftrightarrow \mathrm{PDW}}\sim\sum_{A\neq B}G^{AB}_RG^{AB}_L
\end{equation}
where $A,B=1,2,6,7,8$, describing a SO(5) GN model. The remaining three Majorana fermions $\eta_{3,4,5}$ do not appear in the Hamiltonian so comprise a gapless SO(3) Wess-Zumino-Witten (WZW) model.~\cite{Lin98} Therefore the phase transition between the D-Mott and the PDW is SO(5) GN$\times$SO(3) WZW with a central charge $c=3/2$. 
This may also be called a SU(2)$_2$ criticality or a C0S$3/2$ phase.~\cite{Tsuchiizu02} The notation C$m$S$n$ means a phase with $m$ gapless boson charge fields and $n$ gapless boson spin fields while the subscript of SU(2)$_k$ is the $k$-level and is obtained from $c=3k/(2+k)$. 

\section{Doped carbon nanotube}
\label{sec:doped}
\begin{figure}
\begin{center}
\includegraphics[width=7cm]{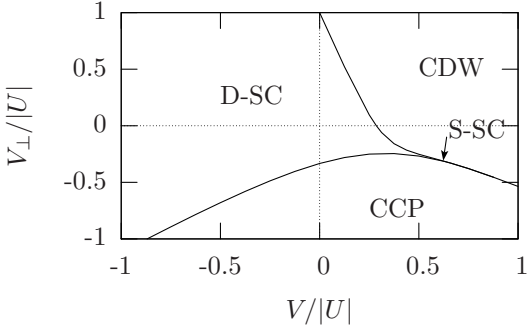}
\caption{Phases for $|V|, |V_{\perp}|<|U|$, $U>0$. The S-SC only exists in a very small region near the junction of the D-SC, CDW and CCP.}
\label{fig:phasedoped+}
\end{center}
\end{figure}

\begin{figure}
\begin{center} 
\includegraphics[width=7cm]{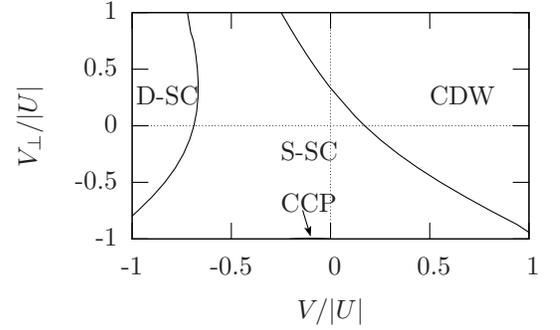}
\caption{Phases for $|V|, |V_{\perp}|<|U|$, $U<0$. There is a very small CCP region for large values of negative $V_{\perp}$ and small values of negative $V$.}
\label{fig:phasedoped-}
\end{center}
\end{figure}

If the CNT is doped we move away from half-filling and $k_F$ deviates from $\pi/2b$ so we can no longer include Umklapp interactions in our Hamiltonian.~\cite{Tsuchiizu05} We assume infinitesimal doping so that the Hamiltonian densities in Eqs. (\ref{eq:Hdens0}) and (\ref{eq:HdensI}) are still valid if all three Umklapp coupling strengths are set to zero. A RG analysis reveals four phases for the doped CNT as shown in Figs. \ref{fig:phasedoped+} and \ref{fig:phasedoped-}, with coupling strengths given in the right column of Table \ref{table:phasesdoped}. 
 Substituting the coupling constants into Eqs. (\ref{eq:Hdens0}) and (\ref{eq:HdensI}) and bosonizing the Hamiltonian gives, for example for the D-SC phase,
\begin{align}
\mathcal{H}_0=&\sfrac{v}{8\pi}\sum_a[(\partial_y\theta_a)^2 +(\partial_y\varphi_a)^2]\nonumber\\
\mathcal{H}_I=&-\frac{g}{2\pi^2}\sum_a\partial_y\phi_{Ra} \partial_y\phi_{La}-4g\sum_{a\neq b}\cos\theta_a\cos\theta_b
\label{eq:dopedH}
\end{align}
for $a,b=2,3,4$ with $\theta_a$ and $\varphi_a$ defined as in Eq. (\ref{eq:DMott}). This new Hamiltonian is similar to Eq. (\ref{eq:DMottH}) for the D-Mott phase, except that $\theta_1=\theta_{\rho_+}$ and $\varphi_1=\varphi_{\rho+}$ do not appear. The pinned values of the D-SC phase and all other observed phases are easily found and given in Table \ref{table:phasesdoped}. 

\begin{table*}
\caption{Pinned fields and coupling strengths of the phases in a doped carbon nanotube.}
\begin{center}
\begin{ruledtabular}
\begin{tabular}{cccccr@{\extracolsep{0pt}\,$>$\,}l}
phase &  $\langle\theta_{\sigma+}\rangle$ & $\langle\theta_{\sigma-}\rangle$ & $\langle\varphi_{\rho-}\rangle$ & $\langle\varphi_{\sigma-}\rangle$ 
& \multicolumn{2}{c}{coupling strength} \\[3pt]
\hline\\
D-SC  & 0 & 0 & 0 & - &
$B_{12}^{\rho}=B^{\sigma}_{12}=2F^{\rho}_{12}=-B_{11}^{\sigma}
=-2B_{11}^{\rho} $&$0$\\
S-SC & 0 & 0 & $\pi$ & -&
$-B_{12}^{\rho}=-B^{\sigma}_{12}=2F^{\rho}_{12}=-B_{11}^{\sigma} =-2B_{11}^{\rho}$&$0$\\
CCP  & 0 & - & 0 & 0&
$-B_{12}^{\rho}=B^{\sigma}_{12}=2F^{\rho}_{12}=-F_{12}^{\sigma} =-2B_{11}^{\rho}$&$0$\\
CDW  & 0 & - & $\pi$ & 0&
$B_{12}^{\rho}=-B^{\sigma}_{12}=2F^{\rho}_{12}=-F_{12}^{\sigma} =-2B_{11}^{\rho}$&$0$\\
\end{tabular}
\end{ruledtabular}
\label{table:phasesdoped}
\end{center}
\end{table*}

We have shown how a D-Mott insulator may be mapped onto a SO(8) GN model. Similarly we can show that a D-SC may be mapped onto a SO(6) GN model of the form 
\begin{equation}
\mathcal{H}=\half\eta_{RA}i\partial_y\eta_{RA}
-\half\eta_{LA}i\partial_y\eta_{LA}
+g G_R^{AB}G_L^{AB}
\end{equation}
with $A,B=3,4\ldots, 8$.
In the undoped case we were able to find mappings from the D-Mott insulator to the seven other phases. Similarly the D-SC can be mapped to the S-SC, the CCP and the CDW in the doped case. For example, the CCP compared to the D-SC is  
\begin{align}
\theta^{CCP}_a&=\left\{\begin{array}{ll}
\theta_a,\qquad& a=2,4\\
\varphi_a & a=3
\end{array}\right.\nonumber\\
\varphi^{CCP}_a&=\left\{\begin{array}{ll}
\varphi_a,\qquad  & a=2,4\\
\theta_a,& a=3.
\end{array}\right.
\end{align}
In terms of the GN Majorana fields
\begin{equation}
\eta^{CCP}_{PA}=\left\{\begin{array}{ll}
P\eta_{PA},\qquad & A=6\\
\eta_{PA},& A=3,4,5,7,8.
\end{array}\right.
\end{equation}
So the D-SC and the CCP share an SO(5) subalgebra. Symmetries between all doped phases can be found in Table \ref{table:symmetries}. 

Essentially one can think of the phases in the doped CNT as being formed from the combination of two of the undoped CNT phases. This is because in the undoped case each phase can be paired with another phase which only differs by the pinned value of $\theta_{\rho+}$, but in the undoped case $\theta_{\rho+}$ is not pinned. For example, the CCP of the doped CNT can be thought of as a combination of the CCP and the FDW of the undoped CNT. Similarly, the CDW of the doped CNT is similar to a combination of CDW and PDW of the undoped CNT and the D(S)-SC is like a combination of the D(S)-Mott and D$'$(S$'$)-Mott phases. 

When we evaluate the order parameters of the doped CNT ground state phases we find that they are comparable to the order parameters of the undoped CNT phases. Using the formulas derived in Sec.  \ref{name-phases} we find that in the CCP of the doped CNT
\begin{align}
j_1(2m)&=i8\sqrt{3}\kappa_{1\uparrow}\kappa_{2\uparrow}
\cos\half\theta_{\rho+}\nonumber\\
j_{11}(2m)&=-i8\sqrt{3}\kappa_{1\uparrow}\kappa_{2\uparrow}
\sin\half\theta_{\rho+}
\end{align}
with $j_1=j_2$ and $j_{11}=j_{12}=-j_{1x}=j_{21}=j_{22}=-j_{2x}$ and all other order parameters vanish. Therefore the CCP of the doped CNT contains both nearest neighbour currents like the CCP of the undoped CNT and next-nearest neighbour currents like the FDW of the undoped CNT. Unlike the undoped case $\theta_{\rho+}$ is not pinned so these currents are not set to some fixed value. Similarly, in the CDW of the doped CNT
\begin{align}
B_1(2m)&=-8\kappa_{1\uparrow}\kappa_{2\uparrow}
\sin\half\theta_{\rho+}
\nonumber\\
B_{11}(2m)&=-8\kappa_{1\uparrow}\kappa_{2\uparrow}
\cos\half\theta_{\rho+}
\end{align}
with $B_1=B_2=B_{\perp}/2$ and $B_{11}=B_{12}=B_{1x}=B_{21}=B_{22}=B_{2x}$
and $n(m)\propto(-1)^m\cos\half\theta_{\rho+}$. All other order parameters are zero in the CDW. In the D-SC and S-SC all  currents and kinetic energies vanish and the deviation from average electron density is zero. The superconducting order parameters in the D-SC are $\Delta_S(m)=\Delta_{D\perp}(2m)=0$ and 
\begin{equation}
\Delta_{D1}(2m)=-\Delta_{D2}(2m)=-2\sqrt{3}
\kappa_{1\uparrow}\kappa_{1\downarrow}e^{i\varphi_{\rho+}/2}
\end{equation}
which is not fixed because $\varphi_{\rho+}$ is not pinned.
In the S-SC, 
\begin{equation}
\Delta_S(m)=4i\kappa_{1\uparrow}\kappa_{1\downarrow}
e^{i\varphi_{\rho+}/2}
\end{equation}
while all the D-SC order parameters are zero.

The phase transitions in the doped nanotube are similar to the transitions in the undoped system. A list of all the transitions and their critical values is given in Table \ref{table:transitionsdoped}. As before, those transitions which require one field's pinned value to change from $0$ to $\pi$ are Gaussian. The massive modes at a Gaussian transition can be shown to be described by a SO(4) GN model. Those transitions in which $\theta_a$ is replaced with its dual field $\varphi_a$ are Ising. The massive modes at the Ising transition are described by an SO(5) GN model. The transitions which are neither Gaussian nor Ising are D-SC$\leftrightarrow$CDW and S-SC$\leftrightarrow$CCP. These two transitions can be understood by substituting their coupling constants into the Hamiltonian density and then mapping to a GN model, as was done for the D-Mott$\leftrightarrow$PDW transition in the undoped case. We find that the Hamiltonian of both transitions resembles an SO(3) GN model, making the transition SO(3) GN$\times$SO(3) WZW, or C0S$3/2$. 

\begin{table*}
\begin{center}
\caption{The phase transitions in the doped case}
\begin{ruledtabular}
\begin{tabular}{ccc}
phase transition & coupling strength & classification
\\[3pt]
\hline\\
D-SC $\leftrightarrow$ S-SC & $B^{\sigma}_{11}=-1$ &
Gaussian\\
CDW $\leftrightarrow$ CCP & $F^{\sigma}_{12}=-1$ &
Gaussian\\
D-SC $\leftrightarrow$ CCP &
$F^{\rho}_{12}=-B^{\rho}_{11}=-F^{\sigma}_{12}=-B^{\sigma}_{11}=1/3$, $B^{\sigma}_{12}=2/3$ & Ising \\
S-SC $\leftrightarrow$ CDW &
$F^{\rho}_{12}=-B^{\rho}_{11}=F^{\sigma}_{12}=B^{\sigma}_{11}=1/3$, $B^{\sigma}_{12}=-2/3$ & Ising \\
D-SC $\leftrightarrow$ CDW & $F^{\rho}_{12}=-B^{\rho}_{11}=1$, $B^{\rho}_{12}=2$
& SO(3) GN$\times$SO(3) WZW\\
S-SC $\leftrightarrow$ CCP & $F^{\rho}_{12}=-B^{\rho}_{11}=1$, $B^{\rho}_{12}=-2$
& SO(3) GN$\times$SO(3) WZW
\end{tabular}
\end{ruledtabular}
\label{table:transitionsdoped}
\end{center}
\end{table*}

\section{Soliton spectrum and quantum numbers}
\label{sec:QN}
In this section we present some general features of the SO($2N$) GN model with special reference to SO(6). Detailed derivations and analyses can be found in Refs. \onlinecite{zamol79,shankar78, karowski80}. The SO($2N$) GN model with integer $N$ is integrable so the soliton excitation spectrum can be calculated exactly. These soliton excitations can be related directly to the CNT by considering quantum numbers such as charge and spin. The SO(8) case has been discussed in Ref. \onlinecite{Lin98}.

\subsection{Excitation spectrum}

A semiclassical analysis of any Hamiltonian of the form given in Eq. (\ref{eq:dopedH}) determines the ground state to be where all $\theta_a=2n_a\pi$ or all $\theta_a=(2n_a+1)\pi$. Therefore we have several possible solutions for the set of all $\theta_a$ for any given ground state. The system can move between any two solutions (in the one ground state) by emitting or absorbing a particle comprised of one or several solitons. The properties of these particles are defined by the changes in $\theta_a$. 

For SO($2N$) $a=1,2,\ldots N$ we consider the change in $\theta_a$ from $y=-\infty$ to $y=\infty$,  
\begin{equation}
\Delta \theta_a=\int_{-\infty}^{\infty}dy \partial_y\theta_a=2\pi N_a.\label{eq:charges}
\end{equation}
where we define the charges $N_a$ as
\begin{equation}
N_a=\int dy\psi_a^{\dag}\psi_a,\qquad a=1,2,\ldots N.
\end{equation}  
The $N$ dimensional vector $(N_1,N_2,\ldots,N_N)$ defines a soliton. 
The simplest solitons have only one $\theta_a$ changing by $2\pi$ over the $y$ range and all the others remaining constant. This corresponds to one $N_a=\pm 1$ and all others zero. These are defined as the fundamental or elementary particles and there will be $2N$ of them. E.g., for SO(6) the fundamental particles are $(\pm 1,0,0)$, $(0,\pm 1,0)$ and $(0,0,\pm 1)$. Another type of soliton changes $\theta_a=2n_a\pi$ to $\theta_a=(2n_a\pm 1)\pi$  for all $a$ over the $y$ range so $N_a=\pm 1/2$ for all $a$. Such a soliton is called a kink. The kinks have no zero charges and each $N_a$ has two possible values and therefore there are $2^N$ kinks. The fundamental fermions and kinks are collectively known as solitons. Thus SO(8) has a total of 24 solitons (16 kinks plus eight fundamental particles), while SO(6) has 14. The even kinks (or simply kinks) are defined to have an even number of positive charges, while the odd kinks (or antikinks) have an odd number of positive charges. For SO(6) the kinks are $(-1,1,1)/2$, $(1,-1,1)/2$, $(-1,-1,-1)/2$, $(1,1,-1)/2$, and the antikinks are $(1,1,1)/2$, $(-1,-1,1)/2$, $(1,-1,-1)/2$ and $(-1,1,-1)/2$. Each fundamental particle may be constructed from a kink-antikink pair if $N$ is even, but if $N$ is odd the fundamental particles are constructed from kink-kink or antikink-antikink pairs. Additional bound states may be constructed from other kink and fundamental particle combinations. 

The masses of all particles constructed from kinks satisfy
\begin{equation}
m_n=2m\sin[n\pi/(2(N-1))],\, n=1,2,\ldots,N-2
\end{equation}
where $m$ is the mass of a kink.
The fundamental particles are $n=1$ and higher values of $n$ describe other bound states. The energy dispersion is given by $\epsilon_n(q)=\sqrt{m_n^2+q^2}$. For $n=N-1$ we obtain $\epsilon_{N-1}(q)=2\sqrt{m^2+q^2/4}\equiv\epsilon_{c}(q)$ above which is a continuum of scattering states. 
In the case of SO(6) the fundamental fermions have mass $m_1=\sqrt{2}m$ which gives $\epsilon_1(q)=\sqrt{2m^2+q^2}$ and there are no higher bound states as  $\epsilon_2=\epsilon_c$. The highest bound states of the SO(8) case have $m_2=\sqrt{3}m$ and are constructed from kink-kink or antikink-antikink pairs. SO(8) is special as $m_1=m$, hinting at the `triality' of this group, i.e, an additional threefold symmetry between kinks, antikinks and fundamental particles. Triality is also indicated by the kinks' and fundamental particles' distance from the origin, being equal to unity for all three of these solitons in SO(8). For SO($2N$) the vectors $(N_1,N_2,\ldots,N_N)$ of the fundamental particles are always a distance of unity from the origin, but for the kinks and antikinks they are a distance of $\sqrt{N}/2$. Therefore for $N\leq 3$ the kinks are closer to the origin than the fundamental particles, but for $N\geq 5$ they are further away. The decreasing distance from the origin of the kink particles as $N$ reduces below 4 implies an increasing instability in the bound states and fundamental fermions. In fact, for $N=2$ there is no fundamental particle spectrum as $\epsilon_1=\epsilon_c$.~\cite{zamol79,shankar78} 

\subsection{Quantum numbers}

Various quantum numbers can be defined by the fermion operators $\psi_{Pj\alpha}$ and may be re-expressed in terms of one of the bosonic fields $\theta_a$ or $\varphi_a$.~\cite{Lin98}
The electronic charge in the two-leg ladder or CNT is defined as
\begin{equation}
Q=\int dy\sum_{Pj\alpha}\psi_{Pj\alpha}^{\dag}\psi_{Pj\alpha}
=\Delta\theta_{\rho+}/\pi.
\end{equation}
The spin is defined by
\begin{equation}
{\bf S}=\int
dy\sum_{Pj\alpha}\psi_{Pj\alpha}^{\dag}(\bs{\sigma}_{\alpha\beta}/2)
\psi_{Pj\alpha}
\end{equation}
from which we obtain
\begin{equation}
S^z=\Delta\theta_{\sigma+}/2\pi.
\end{equation}
The relative $z$-component spin in different bands is
\begin{align}
S^z_{12}&=\int dy\sum_{P\alpha}[\psi_{P1\alpha}^{\dag}\sigma^z_{\alpha\beta}\psi_{P1\alpha}
-\psi_{P2\alpha}^{\dag}\sigma^z_{\alpha\beta}\psi_{P2\alpha}]/2\nonumber\\
&=\Delta\theta_{\sigma-}/2\pi,
\end{align}
while the relative $z$-component vector chirality in different bands is
\begin{align}
P^z_{12}&=\int dy\sum_{P\alpha}P[\psi_{P1\alpha}^{\dag}\sigma^z_{\alpha\beta}\psi_{P1\alpha}
-\psi_{P2\alpha}^{\dag}\sigma^z_{\alpha\beta}\psi_{P2\alpha}]/2\nonumber\\
&=\Delta\varphi_{\sigma-}/2\pi,
\end{align}
and the relative band chirality is
\begin{align}
P_{12}&=\int
dy\sum_{P\alpha}P[\psi_{P1\alpha}^{\dag}\psi_{P1\alpha}-\psi_{P2\alpha}^{\dag}\psi_{P2\alpha}]
\nonumber\\
&=\Delta\varphi_{\rho-}/2\pi.
\end{align}

For all the undoped CNT Mott phases $\Delta\theta_1=\Delta\theta_{\rho+}$, $\Delta\theta_2=\Delta\theta_{\sigma+}$,   
$\Delta\theta_3=\Delta\theta_{\sigma-}$ and 
$\Delta\theta_4=\Delta\varphi_{\rho-}$, which after using Eq. (\ref{eq:charges}) gives $(N_1,N_2,N_3,N_4)=(Q/2,S^z,S^z_{12},P_{12})$. As $\Delta\theta_a=2n_a\pi,(2n_a+1)\pi$ for integral $n_a$ we have $N_a=n_a,n_a+\frac{1}{2}$, showing that the kinks span all possible particles. For the undoped CNT density wave phases only $N_3$ is different when compared to the Mott phases. In this case we have $\Delta\theta_3=\Delta\varphi_{\sigma-}$ so $(N_1,N_2,N_3,N_4)=(Q/2,S^z,P^z_{12},P_{12})$. 

Doping breaks the large SO(8) symmetry but despite this we may still represent the SO(6) solitons as we represented the SO(8) solitons. The SO(6) GN model does not contain $\theta_{\rho+}$ so $Q$ is not a good quantum number and the solitons should be defined by $(S^z,S^z_{12},P_{12})$ in the superconducting sates or $(S^z,P^z_{12},P_{12})$ in the density waves. However, this GN model does not describe the full Hamiltonian. To obtain the full Hamiltonian we must include a chemical potential term to the Hamiltonian, $H=H_{GN}-\mu Q$, where $\mu$ is the chemical potential. Therefore $Q$ is still a good quantum number of the doped Hamiltonian, even though the symmetry between $Q$ and the SO(6) generators is broken.  So, as in the undoped CNT phase, $(N_1,N_2,N_3,N_4)=(Q/2,S^z,S^z_{12},P_{12})$ for the doped CNT Mott phases and 
$(N_1,N_2,N_3,N_4)=(Q/2,S^z,P^z_{12},P_{12})$ for the doped CNT density wave phases.

Other quantum numbers can also be derived, although they will not
be simply related to the $N_a$ charges. For example, the momentum in the $y$ direction (along both legs) may be written as
\begin{align}
P_y=&\int dy\sum_{j\alpha} k_{Fj}(\psi_{Rj\alpha}^{\dag}\psi_{Rj\alpha}
-\psi_{Lj\alpha}^{\dag}\psi_{Lj\alpha})\nonumber\\
=&[(k_{F1}+k_{F2})\Delta\varphi_{\rho+}
+(k_{F1}-k_{F2})\Delta\varphi_{\rho-}]/2\pi.
\end{align}
At half-filling the energy dispersion is degenerate and $k_{F1}=k_{F2}=\pi/2$ so $P_y=\Delta\varphi_{\rho+}/2$. If the system is doped we move away from half-filling so $k_{F1}\neq k_{F2}$. As $\varphi_{\rho+}$ is not defined in the doped case with SO(6) symmetry $P_y$ is not a well defined quantum number.

\section{conclusion}

The metallic zigzag CNT with weak coupling and short-range interactions has a complex phase diagram. We classify all ground states as well as all phase transitions for both the undoped case and the doped case within the parameter range $|V|,|V_{\perp}|<|U|$. We obtain these results by exploiting the relationship between a zigzag CNT and an unusual form of a two-leg ladder. Once the CNT model has been mapped to this two-leg ladder, well established RG and bosonizations techniques are applied to reveal the phase diagram.  

Previous studies on both doped and undoped armchair CNT and their equivalent two-leg ladders have found phases similar to those found here. Though we did expect to find similar phases in these two CNT, we did not expect to find similar phase diagrams as the initial conditions of the RG equations for the armchair and zigzag CNT are very different (Eq. (\ref{eq:initialcoupling}) for the zigzag case). Surprisingly, despite very different initial conditions, the repulsive interactions part of our zigzag CNT phase diagram is remarkably similar to the armchair CNT phase diagram (with interactions $U,V,V_{\perp}>0$ and $t=t_{\perp}$), but with the $V$ and $V_{\perp}$ axes exchanged and the latter axis rescaled.~\cite{Lin02} We find that this is true for both the doped and undoped cases. In other words, the positive $V$ and $V_{\perp}$ segment of Figs. \ref{fig:phase+} and \ref{fig:phasedoped+} is similar to the phase diagram of the armchair CNT with the same interactions and hopping but with $V$ and $V_{\perp}$ exchanged.  This is quite likely related to relative rotations of the graphene lattices in the two CNT as in the zigzag CNT $V$ describes interactions in the longitudinal direction while $V_{\perp}$ describes interactions around the circumference, yet in the armchair CNT $V_{\perp}$ is around the circumference and $V$ is in the longitudinal direction.

Once the ground states have been established a refermionization of their associated Hamiltonians reveal that they can all be mapped onto a SO(2N) GN model, where $N=4$ for the undoped case and $N=3$ for the doped case. In both cases the symmetry is much larger than the symmetry of the initial Hamiltonian so both are examples of dynamical symmetry enlargement. Though all phases in the undoped case have an SO(8) symmetry, they do not all share the same subalgebra. Similarly, in the doped case no two phases share the same SO(6) subalgebra. The subalgebra shared by two phases can hint at the symmetry of the phase transition between these two phases. We observed that if two SO(2N) phases share a SO(2N-1) subalgebra then the phase transition between these two phases is Ising and at the transition the massive fields are defined by a SO(2N-1) GN model. If instead the two phases share a SO(2N-2) subalgebra then the phase transition is Gaussian with the massive fields defined by a SO(2N-2) GN model. In both these cases the fields which are  pinned to equal values on either side of the transition remain massive at the phase transition, while the fields which change their pinned values become massless at the phase transition. This is why the shared symmetry of the two phases is equivalent to the symmetry of the phase transition. We cannot, however, always expect this to be the case. The D-Mott phase and the PDW phase share an SO(3) symmetry and yet at the phase transition the massive modes are described by a SO(5) GN model. This phase transition is distinctly different from the Gaussian and Ising transitions because in this case it is the fields which change their values across the transition which remain massive while the fields which do not change their values becomes massless. 

\section*{ACKNOWLEDGMENTS}
We acknowledge support from the National Science Council of Taiwan through grants NSC-95-2112-M-007-009 and NSC-96-2112-M-007-004 and also support from the National Center for Theoretical Sciences in Taiwan.

%\bibliographystyle{apsrev}
%\bibliography{zigzagnearestn}

\begin{thebibliography}{99}
\expandafter\ifx\csname natexlab\endcsname\relax\def\natexlab#1{#1}\fi
\expandafter\ifx\csname bibnamefont\endcsname\relax
  \def\bibnamefont#1{#1}\fi
\expandafter\ifx\csname bibfnamefont\endcsname\relax
  \def\bibfnamefont#1{#1}\fi
\expandafter\ifx\csname citenamefont\endcsname\relax
  \def\citenamefont#1{#1}\fi
\expandafter\ifx\csname url\endcsname\relax
  \def\url#1{\texttt{#1}}\fi
\expandafter\ifx\csname urlprefix\endcsname\relax\def\urlprefix{URL }\fi
\providecommand{\bibinfo}[2]{#2}
\providecommand{\eprint}[2][]{\url{#2}}

\bibitem[{\citenamefont{Dalton et~al.}(2003)\citenamefont{Dalton, Collins,
  Munoz, Razal, Ebron, Ferraris, Coleman, Kim, and Baughman}}]{Dalton03}
\bibinfo{author}{\bibfnamefont{A.~B.} \bibnamefont{Dalton}},
  \bibinfo{author}{\bibfnamefont{S.}~\bibnamefont{Collins}},
  \bibinfo{author}{\bibfnamefont{E.}~\bibnamefont{Munoz}},
  \bibinfo{author}{\bibfnamefont{J.~M.} \bibnamefont{Razal}},
  \bibinfo{author}{\bibfnamefont{V.~H.} \bibnamefont{Ebron}},
  \bibinfo{author}{\bibfnamefont{J.~P.} \bibnamefont{Ferraris}},
  \bibinfo{author}{\bibfnamefont{J.~M.} \bibnamefont{Coleman}},
  \bibinfo{author}{\bibfnamefont{B.~G.} \bibnamefont{Kim}}, \bibnamefont{and}
  \bibinfo{author}{\bibfnamefont{R.~H.} \bibnamefont{Baughman}},
  \bibinfo{journal}{Nature} \textbf{\bibinfo{volume}{423}},
  \bibinfo{pages}{703} (\bibinfo{year}{2003}).

\bibitem[{\citenamefont{Zhang et~al.}(2005)\citenamefont{Zhang, Fang, Zakhidov,
  Lee, Aliev, Williams, Atkinson, and Baughman}}]{Zhang05}
\bibinfo{author}{\bibfnamefont{M.}~\bibnamefont{Zhang}},
  \bibinfo{author}{\bibfnamefont{S.}~\bibnamefont{Fang}},
  \bibinfo{author}{\bibfnamefont{A.~A.} \bibnamefont{Zakhidov}},
  \bibinfo{author}{\bibfnamefont{S.~B.} \bibnamefont{Lee}},
  \bibinfo{author}{\bibfnamefont{A.~E.} \bibnamefont{Aliev}},
  \bibinfo{author}{\bibfnamefont{C.~D.} \bibnamefont{Williams}},
  \bibinfo{author}{\bibfnamefont{K.~R.} \bibnamefont{Atkinson}},
  \bibnamefont{and} \bibinfo{author}{\bibfnamefont{R.~H.}
  \bibnamefont{Baughman}}, \bibinfo{journal}{Science}
  \textbf{\bibinfo{volume}{309}}, \bibinfo{pages}{1215} (\bibinfo{year}{2005}).

\bibitem[{\citenamefont{Koziol et~al.}(2007)\citenamefont{Koziol, Vilatela,
  Moisala, Motta, Cunniff, Sennett, and Windle}}]{Koziol07}
\bibinfo{author}{\bibfnamefont{K.}~\bibnamefont{Koziol}},
  \bibinfo{author}{\bibfnamefont{J.}~\bibnamefont{Vilatela}},
  \bibinfo{author}{\bibfnamefont{A.}~\bibnamefont{Moisala}},
  \bibinfo{author}{\bibfnamefont{M.}~\bibnamefont{Motta}},
  \bibinfo{author}{\bibfnamefont{P.}~\bibnamefont{Cunniff}},
  \bibinfo{author}{\bibfnamefont{M.}~\bibnamefont{Sennett}}, \bibnamefont{and}
  \bibinfo{author}{\bibfnamefont{A.}~\bibnamefont{Windle}},
  \bibinfo{journal}{Science} \textbf{\bibinfo{volume}{318}},
  \bibinfo{pages}{1892} (\bibinfo{year}{2007}).

\bibitem[{\citenamefont{Zhu et~al.}(2006)\citenamefont{Zhu, Elim, Foo, Yu, Liu,
  Jin, Lee, Shen, Wee, Thong et~al.}}]{Zhu06}
\bibinfo{author}{\bibfnamefont{Y.}~\bibnamefont{Zhu}},
  \bibinfo{author}{\bibfnamefont{I.}~\bibnamefont{Elim}},
  \bibinfo{author}{\bibfnamefont{Y.-L.} \bibnamefont{Foo}},
  \bibinfo{author}{\bibfnamefont{T.}~\bibnamefont{Yu}},
  \bibinfo{author}{\bibfnamefont{Y.}~\bibnamefont{Liu}},
  \bibinfo{author}{\bibfnamefont{W.}~\bibnamefont{Jin}},
  \bibinfo{author}{\bibfnamefont{J.-Y.} \bibnamefont{Lee}},
  \bibinfo{author}{\bibfnamefont{Z.}~\bibnamefont{Shen}},
  \bibinfo{author}{\bibfnamefont{A.~T.-S.} \bibnamefont{Wee}},
  \bibinfo{author}{\bibfnamefont{J.~T.-L.} \bibnamefont{Thong}},
  \bibnamefont{et~al.}, \bibinfo{journal}{Adv. Mater.}
  \textbf{\bibinfo{volume}{18}}, \bibinfo{pages}{587} (\bibinfo{year}{2006}).

\bibitem[{\citenamefont{Yang et~al.}(2008)\citenamefont{Yang, Ci, Bur, Lin, and
  Ajayan}}]{Yang08}
\bibinfo{author}{\bibfnamefont{Z.-P.} \bibnamefont{Yang}},
  \bibinfo{author}{\bibfnamefont{L.}~\bibnamefont{Ci}},
  \bibinfo{author}{\bibfnamefont{J.}~\bibnamefont{Bur}},
  \bibinfo{author}{\bibfnamefont{S.-Y.} \bibnamefont{Lin}}, \bibnamefont{and}
  \bibinfo{author}{\bibfnamefont{P.}~\bibnamefont{Ajayan}},
  \bibinfo{journal}{Nano Lett.} \textbf{\bibinfo{volume}{8}},
  \bibinfo{pages}{446} (\bibinfo{year}{2008}).

\bibitem[{\citenamefont{Tans et~al.}(1998)\citenamefont{Tans, Verschueren, and
  Dekker}}]{Tans98}
\bibinfo{author}{\bibfnamefont{S.~J.} \bibnamefont{Tans}},
  \bibinfo{author}{\bibfnamefont{A.~R.~M.} \bibnamefont{Verschueren}},
  \bibnamefont{and} \bibinfo{author}{\bibfnamefont{C.}~\bibnamefont{Dekker}},
  \bibinfo{journal}{Nature} \textbf{\bibinfo{volume}{393}}, \bibinfo{pages}{49}
  (\bibinfo{year}{1998}).

\bibitem[{\citenamefont{Novak et~al.}(2004)\citenamefont{Novak, Lay, Perkins,
  and Snow}}]{Novak04}
\bibinfo{author}{\bibfnamefont{J.~P.} \bibnamefont{Novak}},
  \bibinfo{author}{\bibfnamefont{M.~D.} \bibnamefont{Lay}},
  \bibinfo{author}{\bibfnamefont{F.~K.} \bibnamefont{Perkins}},
  \bibnamefont{and} \bibinfo{author}{\bibfnamefont{E.~S.} \bibnamefont{Snow}},
  \bibinfo{journal}{Solid-State Electron.} \textbf{\bibinfo{volume}{48}},
  \bibinfo{pages}{1753} (\bibinfo{year}{2004}).

\bibitem[{\citenamefont{Sternberg et~al.}(2006)\citenamefont{Sternberg,
  Curtiss, Gruen, Kedziora, Horner, Redfern, and Zapol}}]{Sternberg06}
\bibinfo{author}{\bibfnamefont{M.}~\bibnamefont{Sternberg}},
  \bibinfo{author}{\bibfnamefont{L.~A.} \bibnamefont{Curtiss}},
  \bibinfo{author}{\bibfnamefont{D.~M.} \bibnamefont{Gruen}},
  \bibinfo{author}{\bibfnamefont{G.}~\bibnamefont{Kedziora}},
  \bibinfo{author}{\bibfnamefont{D.~A.} \bibnamefont{Horner}},
  \bibinfo{author}{\bibfnamefont{P.~C.} \bibnamefont{Redfern}},
  \bibnamefont{and} \bibinfo{author}{\bibfnamefont{P.}~\bibnamefont{Zapol}},
  \bibinfo{journal}{Phys. Rev. Lett.} \textbf{\bibinfo{volume}{96}},
  \bibinfo{pages}{75506} (\bibinfo{year}{2006}).

\bibitem[{\citenamefont{Romero et~al.}(2005)\citenamefont{Romero, Bolton,
  Rosen, and Eklund}}]{Romero05}
\bibinfo{author}{\bibfnamefont{H.~E.} \bibnamefont{Romero}},
  \bibinfo{author}{\bibfnamefont{K.}~\bibnamefont{Bolton}},
  \bibinfo{author}{\bibfnamefont{A.}~\bibnamefont{Rosen}}, \bibnamefont{and}
  \bibinfo{author}{\bibfnamefont{P.~C.} \bibnamefont{Eklund}},
  \bibinfo{journal}{Science} \textbf{\bibinfo{volume}{307}},
  \bibinfo{pages}{89} (\bibinfo{year}{2005}).

\bibitem[{\citenamefont{Dresselhaus et~al.}(2007)\citenamefont{Dresselhaus,
  Dresselhaus, Saito, and Jorio}}]{Dresselhaus07}
\bibinfo{author}{\bibfnamefont{M.~S.} \bibnamefont{Dresselhaus}},
  \bibinfo{author}{\bibfnamefont{G.}~\bibnamefont{Dresselhaus}},
  \bibinfo{author}{\bibfnamefont{R.}~\bibnamefont{Saito}}, \bibnamefont{and}
  \bibinfo{author}{\bibfnamefont{A.}~\bibnamefont{Jorio}},
  \bibinfo{journal}{Annu. Rev. Phys. Chem.} \textbf{\bibinfo{volume}{58}},
  \bibinfo{pages}{719} (\bibinfo{year}{2007}).

\bibitem[{\citenamefont{Meyer et~al.}(2007)\citenamefont{Meyer, Geim,
  Katsnelson, Novoselov, Booth, and Roth}}]{Meyer07}
\bibinfo{author}{\bibfnamefont{J.~C.} \bibnamefont{Meyer}},
  \bibinfo{author}{\bibfnamefont{A.~K.} \bibnamefont{Geim}},
  \bibinfo{author}{\bibfnamefont{M.~I.} \bibnamefont{Katsnelson}},
  \bibinfo{author}{\bibfnamefont{K.~S.} \bibnamefont{Novoselov}},
  \bibinfo{author}{\bibfnamefont{T.~J.} \bibnamefont{Booth}}, \bibnamefont{and}
  \bibinfo{author}{\bibfnamefont{S.}~\bibnamefont{Roth}},
  \bibinfo{journal}{Nature} \textbf{\bibinfo{volume}{446}}, \bibinfo{pages}{63}
  (\bibinfo{year}{2007}).

\bibitem[{\citenamefont{Pan et~al.}(1999)\citenamefont{Pan, Xie, Lu, Chang,
  Sun, Zhou, Wang, and Zhang}}]{Pan99}
\bibinfo{author}{\bibfnamefont{Z.~W.} \bibnamefont{Pan}},
  \bibinfo{author}{\bibfnamefont{S.~S.} \bibnamefont{Xie}},
  \bibinfo{author}{\bibfnamefont{L.}~\bibnamefont{Lu}},
  \bibinfo{author}{\bibfnamefont{B.~H.} \bibnamefont{Chang}},
  \bibinfo{author}{\bibfnamefont{L.~F.} \bibnamefont{Sun}},
  \bibinfo{author}{\bibfnamefont{W.~Y.} \bibnamefont{Zhou}},
  \bibinfo{author}{\bibfnamefont{G.}~\bibnamefont{Wang}}, \bibnamefont{and}
  \bibinfo{author}{\bibfnamefont{D.~L.} \bibnamefont{Zhang}},
  \bibinfo{journal}{Appl. Phys. Lett.} \textbf{\bibinfo{volume}{74}},
  \bibinfo{pages}{3152} (\bibinfo{year}{1999}).

\bibitem[{\citenamefont{Yu et~al.}(2000)\citenamefont{Yu, Lourie, Dyer, Moloni,
  Kelly, and Ruoff}}]{Yu00}
\bibinfo{author}{\bibfnamefont{M.-F.} \bibnamefont{Yu}},
  \bibinfo{author}{\bibfnamefont{O.}~\bibnamefont{Lourie}},
  \bibinfo{author}{\bibfnamefont{M.~J.} \bibnamefont{Dyer}},
  \bibinfo{author}{\bibfnamefont{K.}~\bibnamefont{Moloni}},
  \bibinfo{author}{\bibfnamefont{T.~F.} \bibnamefont{Kelly}}, \bibnamefont{and}
  \bibinfo{author}{\bibfnamefont{R.~S.} \bibnamefont{Ruoff}},
  \bibinfo{journal}{Science} \textbf{\bibinfo{volume}{287}},
  \bibinfo{pages}{637} (\bibinfo{year}{2000}).

\bibitem[{\citenamefont{Yakobson et~al.}(1996)\citenamefont{Yakobson, Brabec,
  and Bernholc}}]{Yakobson96}
\bibinfo{author}{\bibfnamefont{B.~I.} \bibnamefont{Yakobson}},
  \bibinfo{author}{\bibfnamefont{C.~J.} \bibnamefont{Brabec}},
  \bibnamefont{and} \bibinfo{author}{\bibfnamefont{J.}~\bibnamefont{Bernholc}},
  \bibinfo{journal}{Phys. Rev. Lett.} \textbf{\bibinfo{volume}{76}},
  \bibinfo{pages}{2511} (\bibinfo{year}{1996}).

\bibitem[{\citenamefont{Yao et~al.}(2000)\citenamefont{Yao, Kane, and
  Dekker}}]{Yao00}
\bibinfo{author}{\bibfnamefont{Z.}~\bibnamefont{Yao}},
  \bibinfo{author}{\bibfnamefont{C.~L.} \bibnamefont{Kane}}, \bibnamefont{and}
  \bibinfo{author}{\bibfnamefont{C.}~\bibnamefont{Dekker}},
  \bibinfo{journal}{Phys. Rev. Lett.} \textbf{\bibinfo{volume}{84}},
  \bibinfo{pages}{2941} (\bibinfo{year}{2000}).

\bibitem[{\citenamefont{Mintmire et~al.}(1992)\citenamefont{Mintmire, Dunlap,
  and White}}]{Mintmire92}
\bibinfo{author}{\bibfnamefont{J.~W.} \bibnamefont{Mintmire}},
  \bibinfo{author}{\bibfnamefont{B.~I.} \bibnamefont{Dunlap}},
  \bibnamefont{and} \bibinfo{author}{\bibfnamefont{C.~T.} \bibnamefont{White}},
  \bibinfo{journal}{Phys. Rev. Lett.} \textbf{\bibinfo{volume}{68}},
  \bibinfo{pages}{631} (\bibinfo{year}{1992}).

\bibitem[{\citenamefont{Hamada et~al.}(1992)\citenamefont{Hamada, Sawada, and
  Oshiyama}}]{Hamada92}
\bibinfo{author}{\bibfnamefont{N.}~\bibnamefont{Hamada}},
  \bibinfo{author}{\bibfnamefont{S.~I.}~\bibnamefont{Sawada}}, \bibnamefont{and}
  \bibinfo{author}{\bibfnamefont{A.}~\bibnamefont{Oshiyama}},
  \bibinfo{journal}{Phys. Rev. Lett.} \textbf{\bibinfo{volume}{68}},
  \bibinfo{pages}{1579} (\bibinfo{year}{1992}).

\bibitem[{\citenamefont{Saito et~al.}(1992)\citenamefont{Saito, Fujita,
  Dresselhaus, and Dresselhaus}}]{Saito92}
\bibinfo{author}{\bibfnamefont{R.}~\bibnamefont{Saito}},
  \bibinfo{author}{\bibfnamefont{M.}~\bibnamefont{Fujita}},
  \bibinfo{author}{\bibfnamefont{G.}~\bibnamefont{Dresselhaus}},
  \bibnamefont{and} \bibinfo{author}{\bibfnamefont{M.~S.}
  \bibnamefont{Dresselhaus}}, \bibinfo{journal}{Appl. Phys. Lett.}
  \textbf{\bibinfo{volume}{60}}, \bibinfo{pages}{2204} (\bibinfo{year}{1992}).

\bibitem[{\citenamefont{Odom et~al.}(1998)\citenamefont{Odom, Huang, Kim, and
  Lieber}}]{Odom98}
\bibinfo{author}{\bibfnamefont{T.~W.} \bibnamefont{Odom}},
  \bibinfo{author}{\bibfnamefont{J.-L.} \bibnamefont{Huang}},
  \bibinfo{author}{\bibfnamefont{P.}~\bibnamefont{Kim}}, \bibnamefont{and}
  \bibinfo{author}{\bibfnamefont{C.~M.} \bibnamefont{Lieber}},
  \bibinfo{journal}{Nature} \textbf{\bibinfo{volume}{391}}, \bibinfo{pages}{62}
  (\bibinfo{year}{1998}).

\bibitem[{\citenamefont{Wilder et~al.}(1998)\citenamefont{Wilder, Venema,
  Rinzler, Smalley, and Dekker}}]{Wilder98}
\bibinfo{author}{\bibfnamefont{J.~W.~G.} \bibnamefont{Wilder}},
  \bibinfo{author}{\bibfnamefont{L.~C.} \bibnamefont{Venema}},
  \bibinfo{author}{\bibfnamefont{A.~G.} \bibnamefont{Rinzler}},
  \bibinfo{author}{\bibfnamefont{R.~E.} \bibnamefont{Smalley}},
  \bibnamefont{and} \bibinfo{author}{\bibfnamefont{C.}~\bibnamefont{Dekker}},
  \bibinfo{journal}{Nature} \textbf{\bibinfo{volume}{391}}, \bibinfo{pages}{59}
  (\bibinfo{year}{1998}).

\bibitem[{\citenamefont{Gonzalez and Perfetto}(2005)}]{Gonzalez05}
\bibinfo{author}{\bibfnamefont{J.}~\bibnamefont{Gonzalez}} \bibnamefont{and}
  \bibinfo{author}{\bibfnamefont{E.}~\bibnamefont{Perfetto}},
  \bibinfo{journal}{Phys. Rev. B} \textbf{\bibinfo{volume}{72}},
  \bibinfo{pages}{205406} (\bibinfo{year}{2005}).

\bibitem[{\citenamefont{Gonzalez and Perfetto}(2006)}]{Gonzalez06}
\bibinfo{author}{\bibfnamefont{J.}~\bibnamefont{Gonzalez}} \bibnamefont{and}
  \bibinfo{author}{\bibfnamefont{E.}~\bibnamefont{Perfetto}},
  \bibinfo{journal}{Euro. Phys. J. B} \textbf{\bibinfo{volume}{51}},
  \bibinfo{pages}{571} (\bibinfo{year}{2006}).

\bibitem[{\citenamefont{H{\" a}usler et~al.}(2002)\citenamefont{H{\" a}usler,
  Kecke, and MacDonald}}]{Hausler02}
\bibinfo{author}{\bibfnamefont{W.}~\bibnamefont{H{\" a}usler}},
  \bibinfo{author}{\bibfnamefont{L.}~\bibnamefont{Kecke}}, \bibnamefont{and}
  \bibinfo{author}{\bibfnamefont{A.~H.} \bibnamefont{MacDonald}},
  \bibinfo{journal}{Phys. Rev. B} \textbf{\bibinfo{volume}{65}},
  \bibinfo{pages}{85104} (\bibinfo{year}{2002}).

\bibitem[{\citenamefont{Fogler}(2005)}]{Fogler05}
\bibinfo{author}{\bibfnamefont{M.~M.} \bibnamefont{Fogler}},
  \bibinfo{journal}{Phys. Rev. B} \textbf{\bibinfo{volume}{71}},
  \bibinfo{pages}{161304(R)} (\bibinfo{year}{2005}).

\bibitem[{\citenamefont{Kane et~al.}(1997)\citenamefont{Kane, Balents, and
  Fisher}}]{Kane97}
\bibinfo{author}{\bibfnamefont{C.}~\bibnamefont{Kane}},
  \bibinfo{author}{\bibfnamefont{L.}~\bibnamefont{Balents}}, \bibnamefont{and}
  \bibinfo{author}{\bibfnamefont{M.~P.~A.} \bibnamefont{Fisher}},
  \bibinfo{journal}{Phys. Rev. Lett.} \textbf{\bibinfo{volume}{79}},
  \bibinfo{pages}{5086} (\bibinfo{year}{1997}).

\bibitem[{\citenamefont{Egger and Gogolin}(1997)}]{Egger97}
\bibinfo{author}{\bibfnamefont{R.}~\bibnamefont{Egger}} \bibnamefont{and}
  \bibinfo{author}{\bibfnamefont{A.~O.} \bibnamefont{Gogolin}},
  \bibinfo{journal}{Phys. Rev. Lett.} \textbf{\bibinfo{volume}{79}},
  \bibinfo{pages}{5082} (\bibinfo{year}{1997}).

\bibitem[{\citenamefont{Yoshioka and Odintsov}(1999)}]{Yoshioka99}
\bibinfo{author}{\bibfnamefont{H.}~\bibnamefont{Yoshioka}} \bibnamefont{and}
  \bibinfo{author}{\bibfnamefont{A.~A.} \bibnamefont{Odintsov}},
  \bibinfo{journal}{Phys. Rev. Lett.} \textbf{\bibinfo{volume}{82}},
  \bibinfo{pages}{374} (\bibinfo{year}{1999}).

\bibitem[{\citenamefont{Nersesyan and Tsvelik}(2003)}]{Nersesyan03}
\bibinfo{author}{\bibfnamefont{A.~A.} \bibnamefont{Nersesyan}}
  \bibnamefont{and} \bibinfo{author}{\bibfnamefont{A.~M.}
  \bibnamefont{Tsvelik}}, \bibinfo{journal}{Phys. Rev. B}
  \textbf{\bibinfo{volume}{68}}, \bibinfo{pages}{235419}
  (\bibinfo{year}{2003}).

\bibitem[{\citenamefont{Lin et~al.}(1998)\citenamefont{Lin, Balents, and
  Fisher}}]{Lin98}
\bibinfo{author}{\bibfnamefont{H.-H.} \bibnamefont{Lin}},
  \bibinfo{author}{\bibfnamefont{L.}~\bibnamefont{Balents}}, \bibnamefont{and}
  \bibinfo{author}{\bibfnamefont{M.~P.~A.} \bibnamefont{Fisher}},
  \bibinfo{journal}{Phys. Rev. B} \textbf{\bibinfo{volume}{58}},
  \bibinfo{pages}{1794} (\bibinfo{year}{1998}).

\bibitem[{\citenamefont{Konik et~al.}(2000)\citenamefont{Konik, Lesage, Ludwig,
  and Saleur}}]{Konik00}
\bibinfo{author}{\bibfnamefont{R.}~\bibnamefont{Konik}},
  \bibinfo{author}{\bibfnamefont{F.}~\bibnamefont{Lesage}},
  \bibinfo{author}{\bibfnamefont{A.~W.~W.} \bibnamefont{Ludwig}},
  \bibnamefont{and} \bibinfo{author}{\bibfnamefont{H.}~\bibnamefont{Saleur}},
  \bibinfo{journal}{Phys. Rev. B} \textbf{\bibinfo{volume}{61}},
  \bibinfo{pages}{4983} (\bibinfo{year}{2000}).

\bibitem[{\citenamefont{Balents and Fisher}(1997)}]{Balents97}
\bibinfo{author}{\bibfnamefont{L.}~\bibnamefont{Balents}} \bibnamefont{and}
  \bibinfo{author}{\bibfnamefont{M.~P.~A.} \bibnamefont{Fisher}},
  \bibinfo{journal}{Phys. Rev. B} \textbf{\bibinfo{volume}{55}},
  \bibinfo{pages}{R11973} (\bibinfo{year}{1997}).

\bibitem[{\citenamefont{Krotov et~al.}(1997)\citenamefont{Krotov, Lee, and
  Louie}}]{Krotov97}
\bibinfo{author}{\bibfnamefont{Y.~A.} \bibnamefont{Krotov}},
  \bibinfo{author}{\bibfnamefont{D.-H.} \bibnamefont{Lee}}, \bibnamefont{and}
  \bibinfo{author}{\bibfnamefont{S.~G.} \bibnamefont{Louie}},
  \bibinfo{journal}{Phys. Rev. Lett.} \textbf{\bibinfo{volume}{78}},
  \bibinfo{pages}{4245} (\bibinfo{year}{1997}).

\bibitem[{\citenamefont{Lin}(1998)}]{Lin98a}
\bibinfo{author}{\bibfnamefont{H.-H.} \bibnamefont{Lin}},
  \bibinfo{journal}{Phys. Rev. B} \textbf{\bibinfo{volume}{58}},
  \bibinfo{pages}{4963} (\bibinfo{year}{1998}).

\bibitem[{\citenamefont{Tsuchiizu and Furusaki}(2002)}]{Tsuchiizu02}
\bibinfo{author}{\bibfnamefont{M.}~\bibnamefont{Tsuchiizu}} \bibnamefont{and}
  \bibinfo{author}{\bibfnamefont{A.}~\bibnamefont{Furusaki}},
  \bibinfo{journal}{Phys. Rev. B} \textbf{\bibinfo{volume}{66}},
  \bibinfo{pages}{245106} (\bibinfo{year}{2002}).

\bibitem[{\citenamefont{Fjaerestad and Marston}(2002)}]{Fjaerestad02}
\bibinfo{author}{\bibfnamefont{J.~O.} \bibnamefont{Fjaerestad}}
  \bibnamefont{and} \bibinfo{author}{\bibfnamefont{J.~B.}
  \bibnamefont{Marston}}, \bibinfo{journal}{Phys. Rev. B}
  \textbf{\bibinfo{volume}{65}}, \bibinfo{pages}{125106}
  (\bibinfo{year}{2002}).

\bibitem[{\citenamefont{Momoi and Hikihara}(2005)}]{Momoi05}
\bibinfo{author}{\bibfnamefont{T.}~\bibnamefont{Momoi}} \bibnamefont{and}
  \bibinfo{author}{\bibfnamefont{T.}~\bibnamefont{Hikihara}},
  \bibinfo{journal}{J. Phys. Soc. Jpn.} \textbf{\bibinfo{volume}{74}},
  \bibinfo{pages}{226} (\bibinfo{year}{2005}).

\bibitem[{\citenamefont{Tsuchiizu and Suzumura}(2005)}]{Tsuchiizu05}
\bibinfo{author}{\bibfnamefont{M.}~\bibnamefont{Tsuchiizu}} \bibnamefont{and}
  \bibinfo{author}{\bibfnamefont{Y.}~\bibnamefont{Suzumura}},
  \bibinfo{journal}{Phys. Rev. B} \textbf{\bibinfo{volume}{72}},
  \bibinfo{pages}{75121} (\bibinfo{year}{2005}).

\bibitem[{\citenamefont{Bunder and Lin}(2007)}]{Bunder07}
\bibinfo{author}{\bibfnamefont{J.~E.} \bibnamefont{Bunder}} \bibnamefont{and}
  \bibinfo{author}{\bibfnamefont{H.-H.} \bibnamefont{Lin}},
  \bibinfo{journal}{Phys. Rev. B} \textbf{\bibinfo{volume}{75}},
  \bibinfo{pages}{75418} (\bibinfo{year}{2007}).

\bibitem[{\citenamefont{Balents and Fisher}(1996)}]{Balents96}
\bibinfo{author}{\bibfnamefont{L.}~\bibnamefont{Balents}} \bibnamefont{and}
  \bibinfo{author}{\bibfnamefont{M.~P.~A.} \bibnamefont{Fisher}},
  \bibinfo{journal}{Phys. Rev. B} \textbf{\bibinfo{volume}{53}},
  \bibinfo{pages}{12133} (\bibinfo{year}{1996}).

\bibitem[{\citenamefont{Konik et~al.}(2002)\citenamefont{Konik, Saleur, and
  Ludwig}}]{Konik02}
\bibinfo{author}{\bibfnamefont{R.~M.}~\bibnamefont{Konik}},
  \bibinfo{author}{\bibfnamefont{H.}~\bibnamefont{Saleur}}, \bibnamefont{and}
  \bibinfo{author}{\bibfnamefont{A.~W.~W.} \bibnamefont{Ludwig}},
  \bibinfo{journal}{Phys. Rev. B} \textbf{\bibinfo{volume}{66}},
  \bibinfo{pages}{75105} (\bibinfo{year}{2002}).

\bibitem[{\citenamefont{Chen et~al.}(2004)\citenamefont{Chen, Chang, Lin,
  Chang, and Mou}}]{Chen04}
\bibinfo{author}{\bibfnamefont{W.}~\bibnamefont{Chen}},
  \bibinfo{author}{\bibfnamefont{M.-S.} \bibnamefont{Chang}},
  \bibinfo{author}{\bibfnamefont{H.-H.} \bibnamefont{Lin}},
  \bibinfo{author}{\bibfnamefont{D.}~\bibnamefont{Chang}}, \bibnamefont{and}
  \bibinfo{author}{\bibfnamefont{C.-Y.} \bibnamefont{Mou}},
  \bibinfo{journal}{Phys. Rev. B} \textbf{\bibinfo{volume}{70}},
  \bibinfo{pages}{205413} (\bibinfo{year}{2004}).

\bibitem[{\citenamefont{Zamolodchikov and Zamolodchikov}(1979)}]{zamol79}
\bibinfo{author}{\bibfnamefont{A.~B.} \bibnamefont{Zamolodchikov}}
  \bibnamefont{and} \bibinfo{author}{\bibfnamefont{{\Al}.~B.}
  \bibnamefont{Zamolodchikov}}, \bibinfo{journal}{Ann. Phys.}
  \textbf{\bibinfo{volume}{120}}, \bibinfo{pages}{253} (\bibinfo{year}{1979}).

\bibitem[{\citenamefont{Shankar and Witten}(1978)}]{shankar78}
\bibinfo{author}{\bibfnamefont{R.}~\bibnamefont{Shankar}} \bibnamefont{and}
  \bibinfo{author}{\bibfnamefont{E.}~\bibnamefont{Witten}},
  \bibinfo{journal}{Nucl. Phys. B} \textbf{\bibinfo{volume}{141}},
  \bibinfo{pages}{349} (\bibinfo{year}{1978}).

\bibitem[{\citenamefont{Karowski and Thun}(1980)}]{karowski80}
\bibinfo{author}{\bibfnamefont{M.}~\bibnamefont{Karowski}} \bibnamefont{and}
  \bibinfo{author}{\bibfnamefont{H.~J.} \bibnamefont{Thun}},
  \bibinfo{journal}{Nucl. Phys. B} \textbf{\bibinfo{volume}{190}},
  \bibinfo{pages}{61} (\bibinfo{year}{1980}).

\bibitem[{\citenamefont{Lin and Hong}(2002)}]{Lin02}
\bibinfo{author}{\bibfnamefont{H.-H.} \bibnamefont{Lin}} \bibnamefont{and}
  \bibinfo{author}{\bibfnamefont{T.-M.} \bibnamefont{Hong}},
  \bibinfo{journal}{Physica B} \textbf{\bibinfo{volume}{312-313}},
  \bibinfo{pages}{677} (\bibinfo{year}{2002}).

\end{thebibliography}
%\end{document} 
\providecommand{\Yu}{Yu} \providecommand{\Al}{Al} \providecommand{\Ch}{Ch}

\end{document}